# Electron Bloch Oscillations and Electromagnetic Transparency of Semiconductor Superlattices in Multi-Frequency Electric Fields

**Yu. A. Romanov[1], J. Yu. Romanova[1], and L. G. Mourokh[2]**

[1] *Institute for Physics of Microstructures RAS, 603600 Nizhny Novgorod, Russia*

[2]*Department of Physics, Queens College of CUNY, 65-30 Kissena Blvd., Flushing, NY 11365, USA*

## Abstract

We examine phenomenon of electromagnetic transparency in semiconductor superlattices (having various miniband dispersion laws) in the presence of multi-frequency periodic and non-periodic electric fields. Effects of *induced* transparency and spontaneous generation of static fields are discussed. We paid a special attention on a *self-induced* electromagnetic transparency and its correlation to dynamic electron localization. Processes and mechanisms of the transparency formation, collapse, and stabilization in the presence of external fields are studied. In particular, we present the numerical results of the time evolution of the superlattice current in an external *biharmonic* field showing main channels of transparency collapse and its partial stabilization in the case of low electron density superlattices.

## 1. Introduction

Semiconductor superlattices (SLs) are single-crystal composite structures with components periodically changed on the scale (1 ÷ 10 nm) larger than the lattice constant. In such structures, quasimomentum Brillouin zones are broken into a complex of relatively narrow ($10^5 \div 10^7$ cm$^{-1}$) Brillouin *minizones*. In the present time, the mainly studied systems are the SLs with one-dimension periodicity and the electron dispersion law given by

$$\varepsilon(\boldsymbol{k}) = \varepsilon_{||}(k_{||}) + \varepsilon_{\perp}(\boldsymbol{k}_{\perp}), \quad \varepsilon_{\perp}(\boldsymbol{k}_{\perp}) = \frac{\hbar^2 \boldsymbol{k}_{\perp}{}^2}{2m}, \tag{1}$$

where $\varepsilon(\boldsymbol{k})$ and $\boldsymbol{k}$ are the electron energy and quasimomentum, respectively, ($\varepsilon_{||}(k_{||})$, $k_{||}$) and ($\varepsilon_{\perp}(\boldsymbol{k}_{\perp})$, $\boldsymbol{k}_{\perp}$) are their longitudinal and transversal components, respectively, $m$ is the transversal electron effective mass, and $\hbar$ is the Plank constant. We restrict ourselves to the single-miniband approximation, so the number of the miniband is omitted in $\varepsilon(\boldsymbol{k})$ and $\varepsilon_{||}(k_{||})$ of Eq. (1). Electron spin is also neglected. In the present paper, we go beyond the standard nearest-neighbor approximation (sinusoidal miniband). However, to compare our results to previous studies, it is convenient to represent the longitudinal electron energy $\varepsilon_{||}(k_{||})$ as a Fourier sum of *partial* sinusoidal minibands:

$$\varepsilon_{||}(k_{||}) = \sum_{n=1}^{N} \varepsilon_n(k_{||}) = \frac{1}{2}\sum_{n=1}^{N} \Delta_n \left[1 - \cos\left(n k_{||} d\right)\right], \tag{2}$$

where $d$ is the SL period, $\varepsilon_n(k_{||})$ is the dispersion law of the $n$-th partial miniband, $\Delta_n$ is its width, and $N$ is the maximal number of neighbors involved in the dispersion law. It should be emphasized that $n$ is not the *miniband* number (which is omitted) but the number of the *partial* miniband within the given miniband. $\Delta_n$ can be both positive and negative and is determined by the matrix element of the SL Hamiltonian between electron wave functions (Wannier functions) centered on the $\nu$-th and ($\nu$+$n$)-th SL cells (see, e.g. [1]). The full miniband width is $\Delta = \sum_{n=1,3,...} \Delta_n$ . For the nearest-neighbor approximation, $N$ equals to one (sinusoidal dispersion law), and for free electrons one obtains $\Delta_n = (-1)^{n+1} n^{-2} \Delta_1$ and $\Delta = \left(\pi^2/8\right)\Delta_1$ (quasiparabolic dispersion law). Electron dynamics in different partial minibands are not independent because electrons have the same quasimomentum, Brillouin minizone, and quasimomentum distribution function in each of the minibands. The total electron current and energy can be obtained by a summation over all partial currents and energies which are determined by the corresponding Fourier component of the distribution function only. It should be noted that formally electron dynamics in the $\nu$-th partial miniband with period $d$ is equivalent to that of the sinusoidal miniband with the period $\nu d$ and in the Brillouin minizone $\nu$ times narrower than the main one.

Due to the narrowness of Brillouin minizones, Bragg reflections become strongly pronounced in electron dynamics and electrical properties of SLs even at relatively weak electric fields ($10^2 \div 10^4$ V/cm). In particular, at static electric field $E_c$ along the SL axis with negligible interminiband tunneling, these processes lead to a spatial electron localization and create both Bloch oscillations (BOs) [2] (having relatively high Bloch frequency $\Omega_c = eE_c d/\hbar$ and quasiclassic amplitude $Z_c = \Delta/2eE_c$) and a Wannier-Stark ladder of the electron energy levels [3]. This so-called static electron localization occurs in SLs with any miniband dispersion law and in the wide range of the field magnitudes. These magnitudes are limited by electron scattering from the bottom (it is necessary to maintain the condition $\Omega_c \tau_v > 1$, where $\tau_v$ is the electron velocity relaxation time) and by the interminiband tunneling from the top. Rare scattering events do not change the amplitude of BOs, only their phase and center can be changed. The BOs amplitude does not depend on the specifics of the miniband dispersion law which only affects the spectrum of BOs. It should be noted that the BOs are harmonic in the nearest-neighbor approximation only, but in general case, each $\nu$-th partial miniband creates the harmonics with frequency $\nu\Omega_c$. This anharmonicity of BOs plays extremely important role for THz Bloch generators [4].

In the strong time-periodic electric field with fundamental frequency $\omega$ (without the static component), Bragg reflections give rise to nonlinear periodic oscillations with the period of this field and to corresponding quasienergetic minibands [5-11]. We call these oscillations "*ac* BOs" to differ from usual BOs occurring in the static field. Amplitudes of *ac* BOs and quasienergetic minibands widths strongly and non-monotonically depend on the magnitude and spectrum of the field as well as on the miniband dispersion law. In particular, in sinusoidal SL miniband under purely harmonic field, $E_0 \cos(\omega t)$, there are the *ac* BOs with electron velocity having harmonic ($n\omega$) amplitudes $V_{n\omega}=A_n(k_0)J_n(g)$ [5], where $A_n(k_0)$ is a factor depending on the initial electron quasimomentum $k_0$, $J_n(x)$ are the $n$-th order ordinary Bessel functions, and $g = eE_0 d/\hbar\omega$ is the dimensionless field magnitude. Accordingly, at specific values of $g$ determined by the zeros of $J_0(x)$, the electron *dynamic localization* (DL) occurs [5], i.e. the *average* electron velocity is zero independently on initial values of the quasimomentum. In quasienergy representation, it corresponds to the quasienergetic minibands collapse [7] because the finiteness of the motion always leads to the discreteness of the energetic levels. The DL in purely harmonic field is only possible in the SLs with the sinusoidal miniband (i.e. in the nearest-neighbor approximation). In the SLs with more complicated miniband dispersion law, the DL and the miniband collapse can occur in the multi-frequency electric fields only [6,8,10,11]. In particular, if the electric field is a set of rectangular pulses with the magnitude $E_c$ and the period $T_n = 4\pi n/\Omega_c$, ($n = 1,2,..$), it occurs for any arbitrary miniband dispersion law [6,11]. The origin of this is quite obvious because during each half-period an electron experiences the integer number of BOs and returns to the original point. In the SLs with non-sinusoidal miniband given by Eq. (2), the DL can be created by the multi-frequency field $E(t)=\sum_{\alpha=1}^{N} E_\alpha \cos(\omega_\alpha t+\delta_\alpha)$ with non-commensurable frequencies $\omega_\alpha$, arbitrary initial phases $\delta_\alpha$, and amplitudes $E_\alpha$, determined by the conditions $J_0(\alpha g_\alpha) = 0$, where $g_\alpha = eE_\alpha d/\hbar\omega_\alpha$ [8]. It should be noted that in this case *ac* BOs are not periodic and the average electron velocity vanishes (i.e. the DL occurs) only after the averaging over an infinitely large time interval [8].

For the situation when both high-frequency and static components of the electric field are present, electron dynamics is more complicated [12]. It was shown in Refs. [4,8] (see also Eqs. (5) and (29) of the present work) that in the SLs having sinusoidal miniband, the static field shifts the spectra of the *ac* BOs by the Stark frequency $\Omega_c$ with harmonic amplitudes remaining intact. (For clarity, we distinguish positive and negative frequencies, $\pm\, n\omega$.) We will use this property of the *ac* BOs below with the generalization on the SLs with arbitrary miniband dispersion law and arbitrary high-frequency field (not necessary periodic). In particular, at randomly chosen $E_c$, the *ac* BOs are non-periodic and their spectra (for the SLs with sinusoidal miniband) consist of the harmonics with non-commensurable frequencies $\Omega_c \pm n\omega$ only and do not contain zero-th harmonics, i.e. the DL occurs independently on the intensity of the field components (with the electron velocity vanishing only after averaging as in the case discussed above). However, it is evident that for the case of the Stark resonance ($\Omega_c = n\omega$, $n = 1,2,\ldots$), the *ac* BOs in SLs are periodic again. Consequently, the DL and the miniband collapse occur in the SLs with sinusoidal miniband only if the *ac* BOs spectra at $E_c = 0$ do not contain the $n$-th harmonics. If the high-frequency field is harmonic, this happens at amplitudes determined by the zeros of the $n$-th order Bessel functions, $J_n(g) = 0$, [8,12] (see the expression for the electron velocity given above). In the opposite case, electrons are delocalized.

It is evident that *ac* BOs are also periodic when $\Omega_c$ and $\omega$ obey the condition $n_c\Omega_c = n_1\omega$, $n_{c,1}$=1,2... Without loss of generality we can assume that $n_c$ and $n_1$ do not have common integer factor. In this case, the period of *ac* BOs is $T_B = 2n_c\pi/\omega = 2n_1\pi/\Omega_c$ ($n_1/n_c$ is not integer), i.e. this period is larger than the field period in $n_c$ times and larger than the BOs period (at the given static field) in $n_1$ times [8,10,12,13]. Using the picture of the Stark shift of the spectra, one can see that for $n_c > 1$ (!) the *ac* BOs in the SLs with sinusoidal miniband always lack the zero-th harmonics, i.e. the DL always exists. The generalization on the SLs having arbitrary miniband dispersion law was presented in Refs. [10,12,13] (see also Section II of the present paper). It was shown that in such periodic electric field the DL occurs only if there is no partial minibands with $\nu = \mu n_c$, $\mu$ is integer, in the electron dispersion law, independently on the intensity and time-shape of the high-frequency field.

Complex *ac* BOs can manifest themselves *macroscopically* in non-monotonic dependencies of SL nonlinear conductivities upon the magnitude and spectra of its fields. In particular, the effects of *electromagnetic transparency* of harmonic fields in the SLs with the sinusoidal miniband were predicted in Refs. [14,15] and observed experimentally in Ref. [16]. One can distinguish three types of such transparency. Self-induced transparency (SIT) means that at specific magnitudes of high-frequency field the macroscopic electron polarization in the SL becomes small (even vanishes in the limit of rare collisions) and the SL behaves almost like a dielectric with the permittivity of the crystal lattice and relatively small nonlinear absorption. If the SL becomes transparent by means of some applied harmonic field, it can also be transparent for the other harmonic field [9,15,17]. This effect is called induced transparency (IT). Selective transparency (ST) means consequent vanishing of specific harmonics of the excited current with changing of the harmonic field magnitude [8,17]. The mechanisms of the transparency appearing for the SLs with the sinusoidal miniband have been carefully studied in Refs. [8,9]. It was shown that in the single-relaxation-time ($\tau$-) approximation (and in this approximation only!) the DL and the SIT arise at the same harmonic electric fields, despite their different physical origin. The DL occurs immediately after the field application and becomes unobservable in the time $\tau$ due to the *ac* BOs dephasing. A stationary current is established in the SL after the time $\tau$ and its spectral constitution can be significantly different from that of *ac* BOs, especially in the presence of the static component [8,13,18]. In contrast to the DL, the SIT occurs only *after* time $\tau$ from the field application and can exist for a long time (mach larger than $\tau$) [17,19] and, consequently, has more chances to be observed experimentally. Beyond the $\tau$-approximation, the values of the field at which the DL and the SIT occur can be quite different [8,9,17]. Moreover, at non-periodic biharmonic field, the DL exists and the SIT does not [17]. Furthermore, it was shown [17,19] that the transparent state of the SLs with the sinusoidal minband in the harmonic field is unstable with respect to both quasistatic and high-frequency perturbations (in particular, the SL in the SIT state has absolute negative conductivity (ANC) [14,20] and resonant negative dissipative conductivity at frequencies close to the harmonics of the external field [15,17,19]) and, accordingly, can be observed only in the SLs with low electron densities and only in transient processes or under the pulsed external field with the pulse duration much larger than $\tau$.

The influence of weak electron scattering on the DL was studied in details in Ref. [21] for SLs with sinusoidal miniband in harmonic field and in Ref. [13] for arbitrary miniband dispersion and arbitrary fields. In these works, the electric current and the electron diffusion coefficient were determined for various time intervals after the field turn-on. In particular, it was shown that the DL, which occurs at times much smaller than the relaxation time ($t \ll \tau$), is transformed later into qualitatively different state where the time-dependent current vanishes (not just static velocity components of each electron). Authors called both these states (which occur at the same electric fields but in different moments of time within the usual $\tau$-approximation) by the DL. We believe that this term is not suitable for the second state which was predicted in Ref. [14] and called the SIT much earlier than the first publication concerning the DL [5]. The main feature of this state is not the electron spatial localization, i.e. vanishing of their time-averaged velocity and diffusion, but vanishing of time-dependent macroscopic current, i.e. macroscopic SL response. Moreover, the authors' statement that both types of "transparency" occur at the same fields is not always valid even in the $\tau$-approximation. We will show below that it breaks down in the frequency region close to the Bloch frequency and, furthermore, it is not true for the case of non-periodic fields (biharmonic, in particular) without the static component [17].

There are several known scenarios of the SIT collapse for the SL with the sinusoidal miniband in the harmonic field [17,19]. In general case, the SL states in the presence of such field are nonlinear oscillations of the SL current and internal field [17,19,22,23] with the static component present in the spectra. (By the internal field we mean the field which exists in the SL and acts on the SL electrons. This field differs from the external field acting on the SL and depends on the external circuit. This was discussed in details in Refs. [19,22-24].) It could be several possible states resulting in multistability and hysteretic behavior of the dependencies of the generated static field (as well as the current and internal field spectra) on the external field. At high-frequency harmonic external field $E_1\cos(\omega_1 t)$ with $\omega_1\tau \gg 1$, fractionally-quantized static electric fields ($E_c = (\hbar\omega_1/ed)\, p/q$, where $p$ and $q$ are integer) are generated in the SLs with high electron density. However, it was shown in Refs. [17,25] that this generation can be suppressed by the other relatively weak electric field $E_2\cos(\omega_2 t)$ ($g_2 = eE_2 d/\hbar\omega_2 < 1$) with the frequency $\omega_2$ close to $\omega_1$ (but not multiples). In this case, the transparent state of the SL is partially stabilized and, in particular, the ANC disappears. Partial stabilization of the SIT means that the time-dependent current in the SL remains small even with several (usually one or two) harmonics significantly amplified. At low-frequency external fields ($\omega_1\tau \ll 1$), the ANC and the spontaneous generation of the static field do not occur. However, they can arise at *biharmonic* low-frequency field with multiple frequencies ($\omega_2 = n\omega_1$, $\omega_{1,2}\tau \ll 1$, $n = 2,3,\ldots$) at specific values of the amplitudes and phases of its components [25,26].

We believe that for better understanding of nonlinear electromagnetic SL properties, it is necessary to examine their behavior in multi-frequency electric fields beyond the nearest-neighbor approximation (i.e. beyond the sinusoidal miniband dispersion law). Moreover, the internal SL field can contain several frequencies even if the external field is purely harmonic [17,19,22,23]. Our analysis is based on that of Refs. [10,13] (see also references wherein) where the single-electron dynamics are studied in details with arbitrary miniband dispersion law in the presence of arbitrary time-dependent electric field.

In the present work, we study nonlinear oscillations of macroscopic current in semiconductor superlattices with an arbitrary miniband dispersion law in the presence of periodic and non-periodic multi-frequency electric fields. The main attention is paid to the electromagnetic transparency and the processes of its formation, collapse, and stabilization. The rest of the paper is structured as follows. In Section 2, we briefly discuss the main properties of the *ac* BOs and DL of electrons in periodic electric fields with and without a static component. We find numerically the regions with the DL in the biharmonic field with the multiple frequencies and arbitrary component magnitudes. Basic equations for electron transport in SLs in the presence of strong alternating fields are presented in Section 3. Conditions for the SIT formation are determined and its connection to the DL is established. Based on these results, we study the current oscillations in the given *internal* high-frequency biharmonic field in Section 4. Regions of the SIT and ANC are determined numerically. In Section 5, we discuss current-voltage characteristics, nonlinear high-frequency conductivities, and the interaction between electric field components with non-commesurable frequencies in the SLs with an arbitrary miniband dispersion law. In Section 6, we study the peculiarities of interactions between the fields with multiple frequencies in the SLs, formulate the conditions for the spontaneous generation of the static fields, and obtain the equation for the static field magnitude. The current oscillations in the SLs in the presence of given periodic external fields are examined in Section 7 and the processes of the SIT formation, collapse, and stabilization are analyzed. Section 8 contains the formulation of the main results of this work.

## 2. *ac* Bloch Oscillations and dynamical localization of electrons in the superlattice in the presence of multi-frequency fields

Let us start from the brief discussion of the known properties (see Refs. [4,10,13] and references wherein) of the *ac* BOs and DL of electrons in the SLs having arbitrary miniband dispersion law, Eq. (2), in the time-dependent electric field given by

$$E(t) = E_c + \sum_{\alpha=1}^{N} E_\alpha \cos(\omega_\alpha t - \delta_\alpha) , \qquad (3)$$

with the constant component $E_c$ and $N$ harmonics with arbitrary amplitudes $E_\alpha$, frequencies $\omega_\alpha$, and initial phases $\delta_\alpha$. We start from the SLs with the sinusoidal miniband and thereafter generalize the results on SLs with arbitrary dispersion law, Eq.(2). In this, in the contrast with majority of relevant works, where the quantum description of electron dynamics is used, we apply the simple quasiclassic approach because all effects studied in this paper have quaistatic character (except for the postulated dispersion law). We assume that the electric field is always directed along the SL axis and omit the «$_{||}$» subscript in the expressions for the longitudinal energy and longitudinal quasiwavevector.

According to the equation of motion $\dot{k} = eE(t)/\hbar$, an electron in the field of Eq. (3) experiences oscillations with the velocity given by

$$V(k_0,t_0,t) = V_m \sin\left[k_0 d + \Omega_c(t-t_0) + \sum_{\alpha=1}^{N} g_\alpha(\sin(\omega_\alpha t - \delta_\alpha) - \sin(\omega_\alpha t_0 - \delta_\alpha))\right]$$
$$= V_m\left[C_S(k_0,t_0)\Psi_S(t) + C_A(k_0,t_0)\Psi_A(t)\right], \tag{4}$$

where $V_m=\Delta d/2\hbar$ is the maximal longitudinal electron velocity, $k_0$ is its quasiwavevector at the moment $t_0$,

$$\Psi_S(t) = \cos\left[\Omega_c t + \sum_{\alpha=1}^{N} g_\alpha \sin(\omega_\alpha t - \delta_\alpha)\right], \qquad \Psi_A(t) = \sin\left[\Omega_c t + \sum_{\alpha=1}^{N} g_\alpha \sin(\omega_\alpha t - \delta_\alpha)\right] \tag{5}$$

are the wave packets of the electron velocity having different symmetry and spectral components independently on the initial conditions, and

$$C_S(k_0,t_0) = \sin(k_0 d)\Psi_S(t_0) - \cos(k_0 d)\Psi_A(t_0), \;\; C_A(k_0,t_0) = \cos(k_0 d)\Psi_S(t_0) + \sin(k_0 d)\Psi_A(t_0) \tag{6}$$

are the amplitudes of these packets determined by the initial conditions. These amplitudes are changed by electron scattering leading to the energy redistribution between the wave packets without changes in their structures. It should be noted that the following relations remain valid:

$$C_S^2(k_0,t_0) + C_A^2(k_0,t_0) = 1, \;\; \Psi_S^2(t) + \Psi_A^2(t) = 1. \tag{7}$$

It follows from Eqs. (4-6) that the absence of the $\nu$-th harmonics in the spectra of the wave packets $\Psi_{S,A}(t)$ leads to its absence in the electron oscillations spectrum, independently on the initial value of the quasiwavevector. Consequently, the absence of zero-th harmonics in $\Psi_{S,A}(t)$ is the condition for the DL formation. It also follows from Eq. (5) (see also Eq. (29)) that the static field shifts the initial *ac* BO spectrum (spectrum of the electron velocity oscillations in the electric field without the static component) as a whole by $\pm\Omega_c$ with harmonic amplitudes remaining intact (Stark shift) [4,8] which has been mentioned in the Introduction for the harmonic fields. Consequently, in the presence of the static field, the DL occurs in the SLs with the sinusoidal miniband, if the *ac* BO spectrum *without this field* does not contain the harmonics with the $\Omega_c$ frequency.

The generalization on the SLs with arbitrary dispersion law, Eq. (2), is quite obvious. Each $n$-th partial miniband creates $n$-th partial *ac* BO which provides the additive contribution to the complete *ac* BO. Each *n*-th partial *ac* BO is described by Eqs. (4-7) with the replacements $\Omega_c \rightarrow n\Omega_c$ and $g_\alpha \rightarrow ng_\alpha$. The *n*-th partial DL occurs in the SL if in the spectrum of the corresponding oscillations ($\Psi_{S,A}(t)$) the zero-th harmonics is absent. The static field shifts the spectrum of the *n*-th partial *ac* BO by $n\Omega_c$. Consequently, in the presence of the static field the *n*-th partial DL occurs if the *n*-th partial *ac* BO in the field without the static component does not contain the harmonics with the frequency $n\Omega_c$. The DL is complete if it occurs for *all* partial minibands involved in the miniband dispersion law.

In particular, let us assume that the field contains the static and the biharmonic components ($N = 2$ in Eq. (3)). Using the expansion

$$\exp(\pm ig\sin x) = \sum_{n=-\infty}^{\infty} J_n(g)\exp(\pm inx) \ , \tag{8}$$

we obtain from Eq. (5) for this case that

$$\Psi_S(t) = \sum_{\nu_{1,2}=-\infty}^{\infty} J_{\nu_1}(g_1)J_{\nu_2}(g_2)\cos[\nu_1(\omega_1 t - \delta_1) + \nu_2(\omega_2 t - \delta_2) + \Omega_c t],$$
$$\Psi_A(t) = \sum_{\nu_{1,2}=-\infty}^{\infty} J_{\nu_1}(g_1)J_{\nu_2}(g_2)\sin[\nu_1(\omega_1 t - \delta_1) + \nu_2(\omega_2 t - \delta_2) + \Omega_c t]. \tag{9}$$

If the biharmonic field is not periodic ($n_1\omega_1 \neq n_2\omega_2$, $n_{1,2}$ are integers not having the common integer factor, $\omega_1$ and $\omega_2$ are non-commensurable), then the corresponding *ac* BOs are not periodic even at $E_c = 0$. In this field, the DL is formed in the SLs with the sinusoidal miniband at the harmonic amplitudes obeying the condition $J_0(g_1)J_0(g_2) = 0$ for any initial phases but only after infinitely large period of time. In the presence of the static field with $\Omega_c = n_1\omega_1 + n_2\omega_2$, $n_{1,2}$ are integer, positive or negative, including zero, this DL is formed when $J_{n_1}(g_1)J_{n_2}(g_2) = 0$. So, the DL at the harmonic field does not vanish with an addition of the second harmonic field having the non-commensurable frequency. However, the character of its manifestations is changed. In particular, it becomes pronounced after the averaging over a long time interval. If the Bloch frequency does not coincide with significant combinational frequencies ($\Omega_c \neq n_1\omega_1 + n_2\omega_2$), than the DL in the SLs with the sinusoidal miniband occurs for any magnitudes of the field components. (However, this is not observable experimentally because of electron scattering events if $n_{1,2} >> 1$.) For the SLs with arbitrary miniband dispersion law the *n*-th partial DL occurs in the non-periodic biharmonic field with arbitrary amplitudes $E_{1,2}$, if the corresponding *partial* Bloch frequency does not coincide with significant combinational frequencies, $n\Omega_c \neq n_1\omega_1 + n_2\omega_2$. The DL is complete if this inequality is satisfied for all partial minibands involved in the miniband dispersion law.

When

$$n_1\omega_1 = n_2\omega_2 \tag{10}$$

($n_{1,2}$ are integers not having common integer factor), so the frequencies are commensurable, the biharmonic field is periodic with the fundamental frequency

$$\omega_{12} \equiv 2\pi/T_{12} = \omega_1/n_2 = \omega_2/n_1 \tag{11}$$

and the period $T_{12}$. At $E_c = 0$, the *ac* BOs have the same period in the presence of this field and, according to Eq. (9), the DL occurs at magnitudes $E_{1,2}$ and phases $\delta_{1,2}$ determined by the following equations:

$$\begin{aligned} \overline{\Psi}_S(t) &= \sum_{\nu=-\infty}^{\infty} \cos[\nu(n_1\delta_1 - n_2\delta_2)] J_{-n_1\nu}(g_1) J_{n_2\nu}(g_2) = 0, \\ \overline{\Psi}_A(t) &= \sum_{\nu=-\infty}^{\infty} \sin[\nu(n_1\delta_1 - n_2\delta_2)] J_{-n_1\nu}(g_1) J_{n_2\nu}(g_2) = 0. \end{aligned} \tag{12}$$

Here, the bar means the averaging over the field period. For odd values of $n_{1,2}$ (they cannot be even in the same time), the second equation in (12) is satisfied automatically for any $\delta_{1,2}$ and $g_{1,2}$, as well as for the phase shifts given by $\delta \equiv n_1\delta_1 - n_2\delta_2 = \alpha\pi$ ($\alpha = 0,\pm1,\pm2,...$) it is satisfied for any $n_{1,2}$ and $g_{1,2}$. Accordingly, in these two situations the DL occurs in the fields described by the family of *curves* in the $g_1$-$g_2$ plane (or surfaces in the $g_1$-$g_2$-$\delta$ space), in all other cases it happens only at certain *points* in this plane (or curves in the $g_1$-$g_2$-$\delta$ space). To illustrate that, we plot in Fig. 1 the curves for the DL occurrence in the $g_1$-$g_2$ plane at $\omega_2 = 3\omega_1$, $\delta = \pi/2$ and $\pi$, for the SL with sinusoidal miniband. The regions with partial DL with $\nu \neq 1$ can be obtained by the replacement $g_{1,2} \to \nu g_{1,2}$, i.e. by changing of the axes scales by $\nu$ times (regions of partial DL are squeezed and shifted to smaller fields for $\nu > 1$). Complete DL occurs in the SLs at the intersections of partial regions for all ν existing at given miniband dispersion law. They are shown in Fig. 1 by crosses for the case of two partial minibands with $\nu = 1$ and $\nu = 2$ in the biharmonic field with $\omega_2 = 3\omega_1$ and $\delta = \pi/2$. In the three-dimensional $g_1$-$g_2$-$\delta$ space the regions with the complete DL are given by the curves. For the SL having the miniband dispersion law with three partial minibands, the complete DL in the biharmonic field with odd $n_{1,2}$ can occur only at discrete values of $g_1$, $g_2$, and $\delta$. *ac* BOs in the periodic biharmonic field containing the fundamental and even harmonics were analyzed in Refs. [17,27] with account of electron scattering.

In general case, the complete DL in the SL having the miniband dispersion law with $N$ partial minibands can occur only if the field contains not less than $N$ harmonics. However, for some specific case the number of harmonics can be smaller. In particular, according to Eqs. (5,8) in the periodic field containing $M$ harmonics with $E_c = 0$ and frequencies $\omega_\alpha = n_\alpha\omega_1$ ($n_\alpha$ are different integer numbers), the DL regions (and as shown below the SIT in the SL with weak dissipation as well) are given by the solution of the set of $2N$ equations:

$$\begin{aligned} &\sum_{\mu_{2...M}=-\infty}^{\infty} \cos\left(\sum_{\beta=2}^{M} \mu_\beta \delta_\beta\right) J_{-\sum_{\alpha=2}^{M}\mu_\alpha n_\alpha}(\nu g_1) \prod_{\alpha=2}^{M} J_{\mu_\alpha}(\nu g_\alpha) = 0, \\ &\sum_{\mu_{2...M}=-\infty}^{\infty} \sin\left(\sum_{\beta=2}^{M} \mu_\beta \delta_\beta\right) J_{-\sum_{\alpha=2}^{M}\mu_\alpha n_\alpha}(\nu g_1) \prod_{\alpha=2}^{M} J_{\mu_\alpha}(\nu g_\alpha) = 0, \nu = 1...N \end{aligned} . \quad (13)$$

If all $n_\alpha$ are odd, then last N equations are satisfied for all amplitudes $E_\alpha$ and phases $\delta_\alpha$. Correspondingly, in this case the DL occurs at $M \geq (N+1)/2$.

If the periodic field with frequency $\omega$ (in particular, the biharmonic field with frequencies connected by Eq. (10)) also contains the static component which is satisfied the relation

$$\lambda\Omega_c = n\omega, \quad (14)$$

where $n$ and $\lambda$ are the integers not having a common integer multiplier, then *ac* BOs are periodic but their period $T_B$ becomes $\lambda$ times larger than the field period and $n$ times larger than the period of corresponding BOs ($T_B = 2\pi\lambda/\omega = 2\pi n/\Omega_c$). Accordingly, their fundamental frequency is given by

$$\omega_B = \Omega_c/n = \omega/\lambda . \quad (15)$$

To determine the conditions for the DL occurrence, we use the conception of the Stark shift of the *ac* BOs spectra mentioned above. According to this, the zero-th harmonics (leading to the electron delocalization) can appear in the resulting *ac* BOs only in the case when there is at least one $\nu$-th partial miniband in the electron dispersion law for which the Stark shift (which is equal to $\nu\Omega_c$) is the field frequency multiplier. In our case, it is $\nu\Omega_c = \nu\omega n/\lambda$. Correspondingly, the DL occurs in the SL in the periodic field for any arbitrary values of its harmonic amplitudes (the static field value is given by Eq. (14)) if there is no partial minibands with $\nu = \mu\lambda$, where $\mu$ is integer. In particular, it occurs for $\lambda > N$, where $N$ is the maximal partial miniband number in the given miniband. (We do not discuss here accidental compensation of the zero-th velocity harmonics in several partial minibands.) This simple quasiclassic result is in agreement with comprehensive quantum mechanical calculations [10].

At $\lambda = 1$, the field and the *ac* BOs have the same periodicity and the DL in the SLs having sinusoidal miniband occurs only at certain values of the amplitudes and phase of the field harmonics. In particular, in the biharmonic field with frequencies satisfying Eq. (10), it occurs at amplitudes $E_{1,2}$ and phases $\delta_{1,2}$ given by

$$\sum_{\nu=-\infty}^{\infty} \cos[\nu(n_1\delta_1 - n_2\delta_2)] J_{\nu_1^{(0)}-n_1\nu}(g_1) J_{\nu_2^{(0)}+n_2\nu}(g_2) = 0 ,$$

$$\sum_{\nu=-\infty}^{\infty} \sin[\nu(n_1\delta_1 - n_2\delta_2)] J_{\nu_1^{(0)}-n_1\nu}(g_1) J_{\nu_2^{(0)}+n_2\nu}(g_2) = 0, \tag{16}$$

where $\nu_{1,2}^{(0)}$ is the arbitrary pair of integer numbers satisfying

$$n_2\nu_1^{(0)} + n_1\nu_2^{(0)} + n = 0. \tag{17}$$

In particular, for $\omega_2 = n_0\omega_1$, $\Omega_c = n\omega_1$ ($n_0$=2,3,..., $n$ = 1,2,...), we obtain $\nu_1^{(0)} = -n$ and $\nu_2^{(0)}$=0. Similar to the case discussed above, at the phase shift $n_1\delta_1 - n_2\delta_2 = \alpha\pi$ ($\alpha$ = 0,±1,±2,...), the DL occurs at the fields given by the family of curves at the $g_1$-$g_2$ plane and by the family of points at all other shifts.

If the Bloch frequency is non-commensurable with the field frequency ($\lambda\Omega_c \neq n\omega$), then the *ac* BOs are not periodic and their spectra contain only the harmonics having non-commensurable frequencies $\lambda\Omega_c \pm n\omega$ and do not contain the zero-th harmonics. Accordingly, the complete DL in SLs with arbitrary minibands occurs in such field independently on its harmonics intensities and phase values.

## 3. Superlattice transparency. Basic relations

Let us determine the current in the SLs having miniband dispersion law of Eq. (2) in the multi-frequency internal field of Eq. (3). We start from the Boltzmann equation in the $\tau$-approximation given by

$$\frac{\partial f(\boldsymbol{k},t)}{\partial t} + \frac{e\vec{E}(t)}{\hbar}\cdot\frac{\partial f(\boldsymbol{k},t)}{\partial \boldsymbol{k}} = -\frac{f(\boldsymbol{k},t) - f_0(\boldsymbol{k})}{\tau}, \quad f(\boldsymbol{k},t_0) = \tilde{f}_0(\boldsymbol{k}), \tag{18}$$

where $f(\boldsymbol{k},t)$, $f_0(\boldsymbol{k})$ and $\tilde{f}_0(\boldsymbol{k})$ are the field-perturbed, equilibrium, and initial electron distribution functions, respectively, and $t_0$ is the moment of the field turn-on or the moment of the electrons excitation to the conduction band (for instance, by the femtosecond laser). The solution of Eq. (18) in the representation of periodic minibands has the form

$$f(\boldsymbol{k},t) = \tilde{f}_0(\boldsymbol{k} - \Delta\boldsymbol{k}(t,t_0))\exp(-\frac{t-t_0}{\tau}) + \int_{t_0}^{t}\frac{dt'}{\tau}\exp(-\frac{t-t'}{\tau})f_0(\boldsymbol{k} - \Delta\boldsymbol{k}(t,t'))$$
$$\Delta\boldsymbol{k}(t,t')) = \frac{e}{\hbar}\int_{t'}^{t}\boldsymbol{E}(t'')dt'' \tag{19}$$

In general, Eq. (19) allows us to obtain the expressions for the current and electron energies at arbitrary electric fields. However, to get a better understanding of the processes and to simplify the analysis of the self-consistent SLs behavior at external fields, we perform a series of transformations. For simplicity, we assume for a moment that $f_0(\boldsymbol{k}) = \tilde{f}_0(\boldsymbol{k})$. The case of arbitrary initial conditions can be obtained just by adding the exponentially decaying term to the distribution function:

$$\delta f(\boldsymbol{k},t) = \left[\tilde{f}_0(\boldsymbol{k} - \Delta\boldsymbol{k}(t,t_0)) - f_0(\boldsymbol{k} - \Delta\boldsymbol{k}(t,t_0))\right]\exp(-\frac{t-t_0}{\tau}) \tag{20}$$

and corresponding terms in the current and energies. Using the periodicity in the **k**-space, we represent the distribution function in the form

$$f(\boldsymbol{k},t) = \sum_{\nu=-\infty}^{\infty} F_\nu(\boldsymbol{k}_\perp)\exp(i\nu k_3 d)\Phi_\nu(t)\,,\Phi_\nu(t) = a_\nu(t)\Psi_\nu(t) \tag{21}$$

where

$$F_\nu(\boldsymbol{k}_\perp) = \frac{d}{2\pi}\int_{-\pi/d}^{\pi/d} f_0(\boldsymbol{k})\exp(-i\nu k_3 d)dk_3 \quad , \tag{22}$$

$$\Psi_\nu(t) = \exp\left(-i\nu\int_{t_0}^{t}\Omega(t_1)dt_1\right) = \left[\Psi_s(t) - i\Psi_A(t)\right]^\nu \exp\left(i\nu\varphi_0\right) \tag{23}$$

are the eigenfunctions of *ac* BOs. If Eq. (18) is rewritten in term of the complex distribution function $\Phi_\nu(t)$, as

$$\tau\frac{d\Phi_\nu(t)}{dt} + \left[1 + i\nu\tau\Omega(t)\right]\Phi_\nu(t) = 1\,, \tag{18'}$$

with the simplified initial condition $\Phi_\nu(t_0) = 1$, the functions $\Psi_\nu(t)$ are the solutions of Eq. (18') without the collision integral. These functions describe the dynamic (collisionless) modulation of the electron distribution function by the field. Here, functions $\Psi_{S,A}$ are given by Eq. (5), $\varphi_0$ is the real phase given by the field at the initial moment of time, $t_0$, and $a_\nu(t)$ is the dissipative complex function describing modification induced by scattering. This representation is possible only at the chosen simplified boundary conditions; otherwise the function $\Phi_\nu(t)$ would contain the $\boldsymbol{k}_\perp$-dependence, which makes the calculations much more difficult. Substituting the field, Eq. (3), into Eq. (23) and using Eq. (8), we obtain for arbitrary multi-frequency field:

$$\begin{aligned}\Psi_\nu(t) &= \exp\left[i\nu\left(\sum_{\alpha=1}^{M} g_\alpha \sin\left(\omega_\alpha t_0 - \delta_\alpha\right) + \Omega_c t_0\right)\right] \\ &\times \sum_{\mu_{1,2,\dots M}=-\infty}^{\infty}\left\{\prod_{\alpha=1}^{M}\left[J_{\mu_\alpha}\left(\nu g_\alpha\right)\right]\exp\left[-i\left(\nu\Omega_c t + \sum_{\beta=1}^{M}\mu_\beta\left(\omega_\beta t - \delta_\beta\right)\right)\right]\right\}\end{aligned} \tag{24}$$

The intraminiband current $j(t)$ and the relative electron heating $B(t)$ can be expressed in terms of the functions $\Phi_\nu(t)$ as

$$j(t) = \sum_{\nu=1}^{N} j_\nu(t), \quad j_\nu(t) = -j_{0\nu}\,\mathrm{Im}\,\Phi_\nu(t) \tag{25}$$

and

$$B(t) \equiv \frac{\varepsilon(t) - \langle\varepsilon\rangle_0}{\frac{\Delta}{2} - \langle\varepsilon\rangle_0} = \frac{1}{\frac{\Delta}{2} - \langle\varepsilon\rangle_0}\sum_{\nu=1}^{N}\left(\frac{\Delta_\nu}{2} - \langle\varepsilon_\nu\rangle_0\right)B_\nu(t), B_\nu(t) \equiv \frac{\varepsilon_\nu(t) - \langle\varepsilon_\nu\rangle_0}{\frac{\Delta_\nu}{2} - \langle\varepsilon_\nu\rangle_0} = \left(1 - \mathrm{Re}\,\Phi_\nu(t)\right), \tag{26}$$

respectively, where $j_\nu(t)$ is the partial current, $\varepsilon(t)$ and $\varepsilon_\nu(t)$ are the total and partial nonequilibrium electron energies, respectively, $\langle\varepsilon\rangle_0$ and $\langle\varepsilon_\nu\rangle_0$ are the equilibrium energies averaged over the distribution function, $B_\nu(t)$ is the relative electron heating in the $\nu$-th partial miniband (it can be positive or negative),

$$j_{0\nu} = \frac{\nu\Delta_\nu ed}{\hbar}\int f_0(\boldsymbol{k})\cos(\nu k_3 d)\frac{d^3\boldsymbol{k}}{(2\pi)^3} = \frac{e\rho\nu d}{\hbar}\left(\frac{\Delta_\nu}{2} - \langle\varepsilon_\nu\rangle_0\right), \tag{27}$$

and $\rho$ is the electron concentration.

The dissipative function $a_\nu(t)$, describing the changes of the amplitude and frequency spectrum of the $\nu$-th component of the distribution function $\Phi_\nu(t)$ by collisions, obeys equation

$$\frac{da_\nu(t)}{dt} + \tau^{-1}a_\nu(t) = \tau^{-1}\Psi_\nu^*(t), \tag{28}$$

with the initial condition $a_\nu(t_0) = 1$. (Without collisions $a_\nu(t) \equiv 1$.) It should be noted that the reformulation of the problem in terms of the complex function $\Phi_\nu(t)$ instead of the real distribution function $f(\boldsymbol{k},t)$ corresponds to the description in terms of the first partial moments of the distribution function: the average velocity (or current) and average electron energy, as can be seen from Eqs. (25,26). This approach was used in Ref. [19] for the SLs with sinusoidal miniband in the presence of time-dependent electric field and in Ref. [28] for the presence of crossed magnetic and time-dependent electric fields. Further reformulation from the description in terms of functions $\Phi_\nu(t)$ to that of functions $a_\nu(t)$ corresponds to transition to the generalized quasimomentum representation, i.e. to a new set of coordinates, $K_0$, oscillating in the quasimomentum (not in spatial!) space with the collisionless electron. Expansion (21) in this space has the form

$$f^{(K_O)}(k_0,\boldsymbol{k}_\perp,t) = \sum_{\nu=-\infty}^{\infty} F_\nu(\boldsymbol{k}_\perp)\exp(i\nu k_0 d)a_\nu(t),\ k_0 = k_3(t) - \frac{e}{\hbar}\int_{t_O}^{t} E(t)dt\,. \tag{29}$$

In the $K_0$ space, each electron is represented by the fixed point $k_0$ and the distribution of these points can be changed only as a result of collision. In our case of rare collisions, these changes are small over the field period but can be accumulated. In the same time, the equilibrium distribution function $f_0^{(K_0)}(k_0,\boldsymbol{k}_\perp,t)$ in the $K_0$ space is modulated by the field (dynamical modulation) and becomes time-dependent as

$$f_0^{(K_0)}(k_0,\boldsymbol{k}_\perp,t) = f_0\left(k_0 + \frac{1}{d}\int_{t_O}^{t}\Omega(t_1)dt_1, \boldsymbol{k}_\perp\right) = \sum_{\nu=-\infty}^{\infty} F_\nu(\boldsymbol{k}_\perp)\exp(i\nu k_0 d)\Psi_\nu^*(t), \tag{30}$$

which is reflected in the structure of Eq. (28). The outgoing term (the second term in the left-hand-side) has the usual relaxation form and the incoming term (the right-hand-side) is the dynamically modulated equilibrium distribution function. The peculiarity of Eq. (28) caused by the $\tau$-approximation is the absence of relaxation mixing between quasimomentum harmonics with different $\nu$. Income to the given $\nu$-th harmonics is possible only from the equilibrium distribution function. Account of this mixing in the SLs with the sinusoidal miniband was discussed in Refs. [8,9] for some specific cases.

The solution of Eq. (28) with the initial condition $a_\nu(t_0) = 1$ has the form

$$a_\nu(t) = \exp\left(-\frac{t-t_0}{\tau}\right) + \int_{t_0}^{t} \exp\left(-\frac{t-t_1}{\tau}\right)\Psi_\nu^*(t_1)\frac{dt_1}{\tau}. \tag{31}$$

The first term in Eq. (33) describes the general decay of the *ac* BOs (i.e. the destruction of the initial coherent ensemble, see Eq. (21)), the second term is responsible for their transformation into a new wave packet by collisions. Substituting Eqs. (24,31) into Eq. (21), we obtain the following expression for the functions $\Phi_\nu(t)$ determining the current and relative electron heating:

$$\Phi_\nu(t) = \left[1 - \sum_{\mu_{1,2,\dots,M}=-\infty}^{\infty} \frac{\prod_{\alpha=1}^{M} J_{\mu_\alpha}(\nu g_\alpha)\exp\{-i[\nu g_\alpha \sin(\omega_\alpha t_0 - \delta_\alpha) - \mu_\alpha(\omega_\alpha t_0 - \delta_\alpha)]\}}{1 + i\left(\nu\Omega_c + \sum_{\alpha=1}^{M}\mu_\alpha\omega_\alpha\right)\tau}\right]$$

$$\times \exp\left(-\frac{t-t_0}{\tau}\right)\Psi_\nu(t) + \sum_{\lambda_{1,2,\dots,M}=-\infty}^{\infty}\ \sum_{\mu_{1,2,\dots,M}=-\infty}^{\infty} \frac{\prod_{\alpha=1}^{M} J_{\mu_\alpha}(\nu g_\alpha) J_{\mu_\alpha+\lambda_\alpha}(\nu g_\alpha)}{1 + i\left(\nu\Omega_c + \sum_{\alpha=1}^{M}\mu_\alpha\omega_\alpha\right)\tau} \exp\left[-i\sum_{\alpha=1}^{M}\lambda_\alpha(\omega_\alpha t - \delta_\alpha)\right].$$

(32)

If $\tilde{f}_0(\mathbf{k}) \neq f_0(\mathbf{k})$, then one should add the exponentially decaying terms (important for transient processes) to expression for the partial currents and energies as

$$\delta j_\nu(t) = \left[(\tilde{j}_{0\nu} - j_{0\nu})\sin\left(\frac{e\nu d}{\hbar}\int_{t_0}^{t} E(t')dt'\right) + ne\tilde{V}_{0\nu}\cos\left(\frac{e\nu d}{\hbar}\int_{t_0}^{t} E(t')dt'\right)\right]\exp\left(-\frac{t-t_0}{\tau}\right), \tag{33}$$

$$\delta\varepsilon_\nu(t) = \left[\langle\tilde{\varepsilon}_\nu\rangle_0 - \langle\varepsilon_\nu\rangle_0\right]\cos\left(\frac{e\nu d}{\hbar}\int_{t_0}^{t} E(t')dt'\right)\exp\left(-\frac{t-t_0}{\tau}\right), \tag{34}$$

where

$$\tilde{j}_{0\nu} = \frac{\nu\Delta_\nu e d}{\hbar}\int \tilde{f}_0(\mathbf{k})\cos(\nu k_\parallel d)\frac{d^3\mathbf{k}}{(2\pi)^3} = \frac{e\rho\nu d}{\hbar}\left(\frac{\Delta_\nu}{2} - \langle\tilde{\varepsilon}_\nu\rangle_0\right), \tag{35}$$

$\tilde{V}_{0\nu}$, $\langle\tilde{\varepsilon}_\nu\rangle_0$ and $\langle\tilde{\varepsilon}_\perp\rangle_0$ are the averaged velocity and longitudinal and transversal electron energies in the initial moment of time $t_0$, respectively, and $\langle\varepsilon_\perp\rangle_0$ is the equilibrium transversal electron energy.

According to the distribution function $a_\nu(t)$, the first term in Eq. (32) describes the transient partial electron current and energy (corresponding time interval $t << \tau$ is called the *dynamical region* in Refs. [13,21]) and it vanishes at $t >> \tau$ because of collisions. The second term determines the steady values of these quantities (the *kinetic region* in terminology of Refs. [13,21]). In the general case, the frequency spectrum of steady current oscillations is different from the spectrum of the *ac* BOs. The most significant differences occur in the presence of the static component in the periodic electric field with $\nu\Omega_c$ not multiple of the fundamental frequency of the field. In this case, the *ac* BOs are not periodic and their spectrum contains only the harmonics with non-commensurable frequencies $\nu\Omega_c \pm n\omega$. However, it follows from Eq. (32) that for the times of the order of $\tau$, the current spectrum is transformed from the *ac* BOs spectrum in this field to the spectrum containing only the field harmonics (i.e. harmonics of the *ac* BOs in the field without the static component). At the Stark resonance, $\Omega_c = n\omega$ (in the periodic biharmonic field of Eq. (10) $\omega = \omega_{12}$), the *ac* BOs are periodic with the field period and the frequency structure (not the amplitudes) of the spectrum corresponds to the current oscillations. As was shown above, the *ac* BOs are periodic in more general case when the frequencies $\Omega_c$ and $\omega$ are just commensurable, i.e. when $n_c\Omega_c = n_1\omega$, $n_c = 2,3...$, $n_1 = 1,2,...$ The period of the *ac* BOs is given in this case by $T_B = 2n_c\pi/\omega$ ($n_1/n_c$ is not integer), i.e. it is larger than the field period by $n_c$ times. However, after the time of the order of $\tau$, this new periodicity is also destroyed and the current has only the harmonics multiple of the field frequency. The reason for such behavior [8,13,18] is following. As was shown in Ref. [8], the Stark shift of the *ac* BOs spectrum can be viewed as the amplitude modulation of the *ac* BOs by the static field (i.e. just by BOs). In this case, the static field is excluded from the eigenfunctions of the *ac* BOs, $\Psi_\nu(t)$, and the equation for the new dissipative function $\tilde{a}_\nu(t)$ has the form

$$\frac{d\tilde{a}_\nu(t)}{dt} + \left(\tau^{-1} + i\nu\Omega_c\right)\tilde{a}_\nu(t) = \tau^{-1}\Psi_\nu^*(t). \quad (36)$$

Because of the absence of the phase in the static field and because of the collision randomness, the amplitude modulation caused by the static field has an arbitrary phase (randomly changed in time and different for different electrons). Correspondingly, the Stark shift vanishes in the *macroscopic* current, the steady value of which contains only the field harmonics. Formally, this behavior is similar to the absence of the cyclotron shift in plasma subject to the constant magnetic and harmonic electric fields. However, it is different from the Josephson junctions, where collisions do not break the coherence of the Cooper pairs (the superconductive and dissipative currents are in parallel, but the ballistic transfer and collisions in the SLs are in the “series” connection). Accordingly, the steady macroscopic current in the Josephson junctions contains the harmonics with combination frequencies.

If in the multi-frequency field (periodic or non-periodic) there is no static component and the *ac* BOs are high-frequency (i.e. their significant harmonics have only the zero-th harmonic and the harmonics with frequencies $\omega >> \tau^{-1}$), then the *ac* BOs are quite robust to collisions and $a_\nu(t)$ consists of two terms, the main one slightly changing over the field period and the fast-oscillating one which is $\omega\tau$ times weaker. Averaging Eq. (28) (or Eq. (31)) over the time interval $\omega^{-1} << \Delta t << \tau$ (in the case of periodic oscillations, the averaging is performed over their period), we obtain

$$a_\nu(t) = \exp\left(-\frac{t-t_0}{\tau}\right) + \left[1-\exp\left(-\frac{t-t_0}{\tau}\right)\right]\overline{\Psi_\nu^*(t)} + O\left(\frac{1}{\omega\tau}\right). \tag{37}$$

Correspondingly, the SL current is given by

$$j(t) = \frac{i}{2}\sum_{\nu=1}^{N} j_{0\nu}\left[\left(1-\overline{\Psi_\nu^*(t)}\right)\exp\left(-\frac{t-t_0}{\tau}\right) + \overline{\Psi_\nu^*(t)}\right]\Psi_\nu(t) + c.c. + O\left(\frac{1}{\omega\tau}\right) \tag{38}$$

and the relative electron heating in the $\nu$-th partial miniband has the form

$$B_\nu(t) \approx 1 - \mathrm{Re}\left\{\left[\left(1-\overline{\Psi_\nu^*(t)}\right)\exp\left(-\frac{t-t_0}{\tau}\right) + \overline{\Psi_\nu^*(t)}\right]\Psi_\nu(t)\right\}. \tag{39}$$

It follows from Eqs. (21), (28), and (37) that if $\overline{\Psi_\nu^*(t)} = 0$ (conditions for the collapse of $\nu$-th partial miniband and the occurrence of the partial DL), then in the average over the field period the relaxation return of electrons to the state $\nu$, i.e. to the component $\delta f_\nu \sim \exp(i\nu kd)$, is absent (in the $\tau$-approximation!) and, accordingly, this component is washed out from the distribution function for the times of the order of $\tau$. This leads to decoherence of the partial *ac* BO ensemble in such a field and, correspondingly (see Eqs. (38), (39)), to vanishing of the partial current (with $(\omega\tau)^{-1}$ accuracy) in all frequencies, i.e. to the partial SIT, and maximal heating ($\Delta_\nu > 0$) or cooling ($\Delta_\nu < 0$) in the $\nu$-th partial miniband. It is evident that in this case the partial electric field absorption (positive or negative depending on the sign of $\Delta_\nu$) is maximal. Because satisfactions of the conditions $\overline{\Psi_\nu^*(t)} = 0$ for all $\nu$ at the same time are impossible in general case, *ac* BOs decoherence is not complete. It can be complete, in particular, in the multi-frequency field of Eq. (3) with non-commensurable frequencies, $E_c = 0$, $M = N$ and $J_0(\nu g_\nu) = 0$, $\nu = 1,2,\ldots$ [8]. In this field, the functions

$$\overline{\Psi_\nu(t)} = \prod_{\alpha=1}^{N} J_0(\nu g_\alpha) \tag{40}$$

vanish for all $\nu$ at the same time. Correspondingly, at these conditions the complete SIT and DL occur in the SLs with miniband dispersion law containing $N$ partial minibands. For the partial SITs and DLs realization (i.e. for the excluding of some partial miniband contributions to the zero-th harmonics of the *ac* BOs and to the macroscopic current), it is possible to use time-dependent fields with one or several harmonics. Corresponding studies can be used for the characterization of the SL minibands.

It also follows from Eq. (38) that at $\overline{\Psi_\nu^*(t)} \neq 0$, the $\nu$-th partial *ac* BOs are transferred to stationary macroscopic current with the accuracy of two wave packets amplitudes, one of which can be zero (in particular, in the harmonic field without the static component, this can be the amplitude of the wave packet with even harmonics). This is connected to the fact that the structure of the wave packets $\Psi_{A,S}(t)$ does not change at rare electron collisions. Consequently, the selective transparence (ST) occurs in the SLs, i.e. some harmonics of the induced current vanish at the certain values of the field amplitudes and phases. In contrast to SIT, these ranges are determined by the *ac* BOs spectra only, independently on the character of relaxation processes. Accordingly, the ST (similar to the DL) occurs immediately after the field turn-on, not after the time of the order of $\tau$, as the SIT. However, the manifestations of the ST can be seen in the macroscopic current, in contrast to the DL, which becomes non-observable after the time of the order of $\tau$.

The presence of the static component in the periodic field leads to significant peculiarities in the electric current. If the static component satisfies Eq. (14) with $\lambda$ >1, then according to Eq. (24) $\overline{\Psi_\nu^*(t)} \equiv 0$ for all partial minibands with $\nu \neq m\lambda$ ($m$=1,2,...). Consequently, the complete DL and SIP occurs in the SLs independently of the phases and amplitudes of the field harmonics (except for the static one which is given) if the $m\lambda$-th partial minibands are absent in the miniband dispersion law. Furthermore, if there is the $\lambda_0$-th partial miniband in the SL dispersion law for which $\lambda_0\Omega_c = n_0\omega$ ($n_0$ = 1,2,..), i.e. the Bloch frequencies of one or several partial minibands coincide with the field harmonics, then $\overline{\Psi_{\lambda_0}(t)} \neq 0$. In this case, the DL and SIP can occur only in periodic fields with specific spectral constitution, in particular in the harmonic field $\overline{\Psi_{\lambda_0}(t)} = J_{n_0}(\lambda_0 g)$. Consequently, if this resonance takes place for the $\lambda_0$-th harmonics only, then the SIT occurs at the field amplitudes satisfying $J_{n_0}(\lambda_0 g) = 0$. In the opposite case, the SIT can be only partial.

The current induced by the multi-frequency field in SLs is small, i.e. $j(t)$ ~$(\omega\tau)^{-1}$, and the SL is in the state of complete SIT in the case when the partial Bloch frequencies do not coincide with significant combinational frequencies of a periodic or non-periodic field, $\left|\lambda\Omega_c - \sum_{\alpha=1}^{M} \nu_\alpha \omega_\alpha\right| >> \tau^{-1}$, $\Omega_c\tau >> 1$ ($\nu_\alpha$ are small integers), and the harmonics amplitudes are arbitrary. As was shown before for the case of the biharmonic field, the complete DL can also occur in this situation.

Peculiarities of the alternating current in SLs in the presence of high-frequency fields described above are qualitatively similar to that of the SLs with the sinusoidal miniband in the harmonic field [8] and in agreement with conclusions of Refs. [13,21]. In particular, in all these cases the field regions for the occurrences of the DL and SIT coincide (in the $\tau$-approximation!). However, it should be emphasized that this is caused by the approximations of Eq. (28) which does not take into account the relaxation mixing of the quasimomentum harmonics with different $\nu$. If the energy dependence of the relaxation time is significant (in particular, for scattering on optical phonons), Eq. (28) is not valid anymore. Consequently, there is no SIT for the fields where the DL occurs. Strictly speaking, the regions for the DL and SIT occurrences are always different. Moreover, in contrast to the DL, the existence of the SIT depends on specific scattering mechanisms and in the situation where the main source of scattering is the interaction with optical phonons, the SIT does not occur at any fields. This is extremely important for experimental studies.

There is important specific case (even in the $\tau$-approximation), when the Stark frequency $\lambda\Omega_c$ of one or several partial minibands is close to one of the combination frequencies of the periodic or non-periodic field, i.e $|\lambda\Omega_c - \sum_{\alpha=1}^{M} \nu_\alpha^0 \omega_\alpha | \sim \tau^{-1}$, where $\{\nu_\alpha^0\}$ is the set of specific integer numbers. In these resonant regions, the *ac* BOs contain strongly decaying low-frequency harmonics (with frequencies of order of the inverse electron relaxation time) and consequently, as follows from Eq. (32), the currents induced by the $\lambda$-th partial miniband can have extremely large values at all frequencies:

$$j_\lambda \sim \frac{i j_{0\lambda} \prod_{\alpha=1}^{M} J_{-\nu_\alpha^0}(\lambda g_\alpha)}{1 + i\left(\lambda\Omega_c - \sum_{\alpha=1}^{M} \nu_\alpha^0 \omega_\alpha\right)\tau} \sum_{\mu_{1,2,\ldots,M}=-\infty}^{\infty} \prod_{\alpha=1}^{M} J_{\mu_\alpha - \nu_\alpha^0}(\lambda g_\alpha) exp\left[-i\sum_{\alpha=1}^{M} \mu_\alpha(\omega_\alpha t - \delta_\alpha)\right] \sim j_{0\lambda} . \quad (41)$$

In other words, in these resonant regions, despite the presence of the complete DL at any amplitudes of the multi-frequency field, the SL transparency occurs only in the absence of at least one of the $\nu_\alpha^0 \omega_\alpha$ harmonics (involved in the resonant induced energy exchange, see Section V) in the corresponding partial *ac* BO (without $E_c$), i.e. at $J_{\nu_\alpha^0}(\lambda g_\alpha) = 0$. Peculiar situation occurs in the non-periodic field without the static component in regions $|\sum_{\alpha=1}^{N} \nu_\alpha \omega_\alpha | \sim \tau^{-1}$ [15,17]. In these fields, the complete DL occurs at the field amplitudes given by the zeros of the Bessel function $J_0(\alpha g_\alpha) = 0$ (see Eq. (40)) and the SIT and ST are absent at all field amplitudes (see Eqs. (32),(41)). (In this case, the term followed from Eq. (32) with $\mu_\alpha = 0$ should be added to the current, Eq. (41) at $\Omega_c = 0$.) The physical mechanisms and consequences of such current behavior will be discussed below.

Thus, if the *ac* BOs do not contain low-frequency harmonics, then in the $\tau$-approximation (and in this approximation only [6]) the SIT of the SLs and the DL occurs in the same fields. However, it does not mean that they are the manifestations of the same phenomenon. The DL is the characteristics of the single-electron dissipationless dynamics with all harmonics to be significant except the zero-th one. The DL occurs right after the electric field turn-on (or right after the electron appearance in the miniband) and becomes unobservable in the macroscopic current after the time of the order of $\tau$ because of decoherence of the *ac* BOs ensemble. The SIT characterizes the collective dissipative electron behavior, manifests itself in *all* current harmonics, develops for the time of the order of $\tau$ after the electric field turn-on (or after the electron appearance in the miniband), and can exist for significant period of time. It is possible to say that in the case of high-frequency *ac* BOs, the DL is the reason (in the $\tau$-approximation) of the SIT which develops by collisions for the time of the order of $\tau$, whereas for the *ac* BOs containing low-frequency harmonics, the DL does not lead to the SIT.

## 4. SL transparence in the periodic biharmonic field

Let us consider the SL behavior in the periodic biharmonic field without the static component, $\omega_2 = n_0\omega_1$, $n_0 = 2,3,\ldots$ , $\delta_1 = 0$, $\delta_2 \equiv \delta$. In this case, according to Eq. (24), the eigenfunction of the *ac* BOs is given by (with accuracy to an unsignificant phase factor)

$$\psi_\nu(t) = \sum_{l,\mu=-\infty}^{\infty} J_{l-n_0\mu}(\nu g_1) J_\mu(\nu g_2) \exp[-i(l\omega_1 t - \mu\delta)] \tag{42}$$

Substituting Eq. (42) into Eq. (38), we obtain the following expression for high-frequency steady current at $\omega_1\tau >> 1$:

$$j(t) = \sum_{\nu=1}^{N} \sum_{l=-\infty}^{\infty} j_{l\omega_1}^{(\nu)} \exp[-il\omega_1 t] + \text{c.c} + O\left(\frac{1}{\omega_1\tau}\right), \tag{43}$$

where

$$j_{l\omega_1}^{(\nu)} = \frac{i}{2} j_{0\nu} \sum_{\lambda=-\infty}^{\infty} J_{l-n_0\lambda}(\nu g_1) J_\lambda(\nu g_2) \sum_{\mu=-\infty}^{\infty} (-1)^{n_0\mu} J_{n_0\mu}(\nu g_1) J_\mu(\nu g_2) \exp[i(\lambda-\mu)\delta]. \tag{44}$$

The internal sum in Eq. (44) coincides with $\exp(i\lambda\delta)\overline{\Psi_\nu^*(t)}$. Equality of this sum to zero (see Eq. (12)) provides the condition for the occurrences of the partial DL (miniband collapse) and the SIT. The regions of the occurrence of the partial SIT with ν = 1 ($\overline{\Psi_1(t)}=0$), i.e. the SIT in the SL having sinusoidal miniband, in the biharmonic electric field with $\omega_2 = 3\omega_1$ and phase shifts $\delta = \pi$ and $\delta = \pi/2$ are shown in Fig.1. The SIT states in the SL containing two partial minibands with $\nu = 1$ and $\nu = 2$ are marked by crosses. In the SL containing three partial minibands, the complete DL and SIT occur in the biharmonic field only with odd harmonics with discrete values of their amplitudes and phases, such as $g_1$ = 1.564, $g_2$ = 2.478, $\delta = 0.472\pi$; $g_1$ = 2.404, $g_2$ = 2.596, $\delta = \pi/2$; and so on. It is evident that for the SLs containing more than three partial minibands, the complete SIT does not occur in the biharmonic field. Let us recall that the conditions for the DL and SIT occurrences coincide only at the $\tau$-approximation [8,9].

It is interesting to examine the case of odd $n_0$ and $\delta = \pi/2$ (Fig. 1(b)). At transparency conditions in the purely harmonic field, given by zeros of the zero-th Bessel function ($J_0(g_1^\lambda)=0$, where $\lambda$ is the number of the Bessel function root), the SL with sinusoidal dispersion law is almost transparent for the second field as well, up to its amplitudes $g_2 \sim g_1$, if $g_1 << 2n_0$, $\lambda < (n_0+1)/2$. This is a manifestation of the high-order Bessel function properties. Indeed, in this case the second equation of Eq. (12) is the identity and the first one has the form

$$J_0(g_1)J_0(g_2)+2\cdot\sum_{\mu=1}^{\infty}(-1)^{\mu}J_{2\mu n_0}(g_1)J_{2\mu}(g_2)=0\ . \qquad (45)$$

The first term in Eq. (45) is equal to zero and the second one is small when $g_1 << 2n_0$, $\lambda < (n_0+1)/2$ because $J_{2\mu n_0}(g_1)$ is small. In particular, for $n_0 = 3$ it happens at $g_1$ = 2.405, for $n_0 = 5$ at $g_1$ = 2.405 and 5.5, and so on. The exceptions are the points near $J_0(g_2) = 0$, because the value of $J_2(g_2)$ is significant there. This phenomenon can be called the *induced transparency* (IT) in the fields with multiple frequencies. In contrast to the IT in the fields with non-multiple frequencies [15], the phase shift between the fields is important here.

## 5. Current-voltage characteristics, nonlinear conductivities, and the field interaction

To further understand the SL behavior at the state of the SIT, let us examine its current-voltage characteristics and nonlinear conductivities. Corresponding studies for the case of the SLs with sinusoidal miniband in the presence of a biharmonic field were performed in Refs. [15,17,19,25,26,29]. In the present work, we present more general results including arbitrary minibands and arbitrary electric fields. If the components of the field of Eq. (3) are non-commensurable (the case of commensurable frequencies is examined in the next section), i.e. $\mu_\alpha\omega_\alpha \neq \mu_\beta\omega_\beta$, then, according to Eqs. (25) and (32), the static current,

$$j_c \;=\; \sum_{\nu=1}^{N}\sigma_c^{(\nu)}(\Omega_c,\Omega_1,\ldots,\Omega_M,\omega_1,\ldots,\omega_M)E_c\ , \qquad (46)$$

and the currents at $\omega_\alpha$-frequencies,

$$j_{\omega_\alpha}(t) = \mathrm{Re}\sum_{\nu=1}^{N}\sigma^{(\nu)}(\omega_\alpha;\Omega_c,\Omega_1,...\Omega_M,\omega_1,\omega_2,...\omega_M)E_\alpha \exp[-i(\omega_\alpha t-\delta_\alpha)], \tag{47}$$

are completely determined by nonlinear partial conductivities (independent on the phase relations between fields):

$$\sigma_c^{(\nu)}(\Omega_c,\Omega_1,...,\Omega_M,\omega_1,...,\omega_M) = \frac{\sigma_{0\nu}}{\nu\Omega_c}\sum_{\mu_{1,2...M}=-\infty}^{\infty}\frac{\nu\Omega_c+\sum_{\alpha=1}^{M}\mu_\alpha\omega_\alpha}{1+\left(\nu\Omega_c+\sum_{\alpha=1}^{M}\mu_\alpha\omega_\alpha\right)^2\tau^2}\prod_{\beta=1}^{M}J_{\mu_\beta}^2(\nu g_\beta) \tag{48}$$

and

$$\sigma^{(\nu)}(\omega_\alpha;\Omega_c,\Omega_1,...,\Omega_M,\omega_1,...,\omega_M)$$

$$= \frac{\sigma_{0\nu}}{(\nu\Omega_\alpha)^2}\sum_{\mu_{1,2,...M}=-\infty}^{\infty}\left\{\left[2\mu_\alpha\omega_\alpha\left(\nu\Omega_C+\sum_{\beta=1}^{M}\mu_\beta\omega_\beta\right)-i\frac{\Omega_\alpha}{\tau}\frac{d}{dg_\alpha}\right]\frac{\prod_{\beta=1}^{M}J_{\mu_\beta}^2(\nu g_\beta)}{1+\left(\nu\Omega_C+\sum_{\beta=1}^{M}\mu_\beta\omega_\beta\right)^2\tau^2}\right\}, \tag{49}$$

where $\sigma_{0\nu} = \frac{e\nu d\tau}{\hbar}j_{0\nu}$ is the partial linear conductivity. In this situation the Joule losses are given by

$$\overline{j(t)E(t)} = \sum_{\nu=1}^{N}j_{0\nu}\frac{\hbar}{e\nu d\tau}\left[1-\sum_{\mu_{1,2...M}=-\infty}^{\infty}\frac{\prod_{\beta=1}^{M}J_{\mu_\beta}^2(\nu g_\beta)}{1+\left(\nu\Omega_C+\sum_{\beta=1}^{M}\mu_\beta\omega_\beta\right)^2\tau^2}\right]. \tag{50}$$

(It should be noted that Eq. (29) of our previous work, Ref. [19], for the dissipative losses in the SLs with the sinusoidal miniband in the harmonic field is invalid. The right expression can be obtained from Eq. (50) of the present paper at $M = N = 1$.) It follows from Eq. (50) that the maximal possible absorption in SL (occurring at strong electron heating by an electric field in the miniband of width $\Delta$) is determined by $Q_m \sim n\Delta/\tau$.

Comparing Eqs. (49) and (50), we obtain the useful expression (in the $\tau$-approximation)

$$\mathrm{Im}\,\sigma(\omega_\alpha;\Omega_C,\Omega_1,...\Omega_M,\omega_1,\omega_2,...\omega_M) = \left(\frac{e\nu d}{\hbar}\right)^2\frac{\tau}{\omega_\alpha}\frac{d}{dg_\alpha}\overline{j(t)E(t)}. \tag{51}$$

One can see from Eqs. (50) and (51) that at rare collisions ($\omega_\alpha\tau >> 1$) the full dissipative loses in SLs are the oscillating functions of the field harmonics frequencies and amplitudes and, consequently, the reactive conductivities are the sign-changing functions of these variables. This means that it is possible to have several branches of plasma oscillations in the SL subject to a given time-dependent field. The excitation of plasma oscillations is the main path for the SIT collapse in SLs [17,19]. In the SLs with low electron density, this collapse leads to the spontaneous generation of the static fields. This statement is valid for the SLs with arbitrary miniband dispersion law in the presence of multi-frequency periodic fields as well. However, the corresponding process of the SIT collapse becomes multi-channel because of several brances of plasma oscillations. Correspondingly, the number of possible final states (to which the SLs with non-sinusoidal miniband can be transformed after the SIT collapse) increases as well.

It is evident from Eqs. (48,49) (see also Eq. (41)) that at $\omega_\alpha\tau >> 1$ nonlinear currents, conductivities, and (nondissipative!) energy loses for field components have resonant singularities in narrow frequency ranges $\left|\nu\Omega_c - \sum_{\alpha=1}^{M}\mu_\alpha^{(0)}\omega_\alpha\right| \approx \tau^{-1}$, $\mu_\alpha^{(0)} = 0, \pm 1, \pm 2, ...$ At these frequencies, there is almost dissipationless sign-changing energy exchange between static and time-dependent fields described by the expressions

$$j_c E_c \approx -\frac{1}{2}\mathrm{Re}\sum_{\alpha=1}^{M}\sigma^{(\nu)}(\omega_\alpha;\Omega_c,\Omega_1,...,\Omega_M,\omega_1,...,\omega_M)E_\alpha^2$$

$$\sim j_{0\nu}E_c\frac{\left(\nu\Omega_c - \sum_{\alpha=1}^{M}\mu_\alpha^{(0)}\omega_\alpha\right)\tau}{1+\left(\nu\Omega_c - \sum_{\alpha=1}^{M}\mu_\alpha^{(0)}\omega_\alpha\right)^2\tau^2}\prod_{\beta=1}^{M}J^2_{\mu_\beta^{(0)}}(\nu g_\beta), \tag{52}$$

$$|j_c E_c| >> \overline{j(t)E(t)} \sim Q_m.$$

This energy exchange is carried out by the nonlinear electron *ac* BOs. According to the energy conservation law, in the $\nu$-th partial channel with $\Delta_\nu > 0$ the energy flow is directed from the field with larger harmonic frequency to the field with the smaller harmonic frequency and for the channel with $\Delta_\nu < 0$, the energy flow direction is reversed. For the static field, the Bloch frequency $\nu\Omega_c$ plays the role of the harmonic frequency. In particular, let us consider the case with $\Delta_\nu > 0$. In this situation, the electromagnetic energy flows from the static field to the high-frequency ones if $\nu\Omega_c > \sum_{\alpha=1}^{M}\mu_\alpha^{(0)}\omega_\alpha$ . In this, the electron is shifted by $\nu$ SL periods against the static field, gains the energy of $e\nu E_c d$, and "emits" $\{\mu_\alpha^{(0)}\}$ quanta of the alternating fields. The rest of the energy ($\sim Q_m$) goes to the lattice. In this kind of multi-photon processes, at $\mu_\alpha^{(0)}>0$ the field quanta are emitted and at $\mu_\alpha^{(0)}<0$ the field quanta are absorbed. Thermal equilibrium state of environment (for example, phonon heat bath) manifests itself in the fact that in average electrons lose their kinetic energy via the interaction with environment. If $\nu\Omega_c < \sum_{\alpha=1}^{M}\mu_\alpha^{(0)}\omega_\alpha$ , the process is reversed and the energy flows from the high-frequency field to the static one, i.e. the electron

“absorbs” $\{\mu_\alpha^{(0)}\}$ quanta of the alternating fields and moves by $v$ SL periods along the field direction. The rest of the energy goes to the lattice as well. In the both cases described above $|j_c E_c| >> \overline{j(t)E(t)}$, i.e. only a small part of the energy goes to the lattice, but mainly the energy is redistributed between the field components. The maximal energy exchange occurs at $v\Omega_c \approx \sum_{\alpha=1}^{M} \mu_\alpha^{(0)}\omega_\alpha \pm \tau^{-1}$. With accuracy of $(\Omega_c\tau)^{-1}$ (see Eq. (52)) it does not depend on $\tau$. Such almost dissipationless energy exchange can happen between different high-frequency components as well (with or without the static component in the field) [15,19]. At $v\Omega_c \approx \sum_{\alpha=1}^{M} \mu_\alpha^{(0)}\omega_\alpha$ the static current $j_c$ changes its direction with respect to the static field $E_c$ (see Eq. (52)), i.e. in part of these regions the SL has the *absolute negative conductivity* (ANC).

Thus, one can see that the occurrence of significant current in the resonant regions is related to the induced energy exchange between the field components. As was shown above, the DL occurs in these regions but the SIT is absent. However, if at least one of the $\mu_\alpha^{(0)}\omega_\alpha$ harmonic is absent in the corresponding *ac* BOs ($J^2_{\mu_\alpha^{(0)}}(vg_\alpha) = 0$), then this energy flow channel vanishes and the SIT occurs, as we showed above. It is evident that in this case the inequality of Eq. (52) is broken.

In general case, the energy exchange between the static and high-frequency fields manifests itself in the oscillatory behavior of current-voltage characteristics and dynamic conductivities of SLs. As an example, in Fig.2 we plot current-voltage characteristics and conductivities $\mathrm{Re}\sigma(\omega_{1,2};\Omega_c,\Omega_1,\Omega_2,\omega_1,\omega_2)$ for the SL with harmonic miniband dispersion law in high-frequency biharmonic field with $\omega_2 = 1.3\omega_1$, $\omega_1\tau = 10$ and $g_1 = g_2 = 2.405$ as functions of $g_c = \Omega_c/\omega_1$. One can see in this figure the above mentioned peculiarities at the Bloch frequency $\Omega_c \approx \omega_{1,2}$, $n_2\omega_2 \pm n_1\omega_1$, $n_{1,2} = \pm1,\pm2,\ldots$ . The shown current-voltage characteristics have the regions both with negative differential conductivity (NDC) and with ANC. The existence of NDC leads to the domain instability, whereas the existence of the region with ANC means the instability of the SIT states with respect to a spontaneous generation of the static field which will be discussed below. According to Eqs. (46-49), the current-voltage characteristics and nonlinear conductivities for the SLs with arbitrary miniband can be obtained by the summation of corresponding partial counterparts, and the latter are determined by the curves of Fig.2 with the replacement $g_c \to vg_c$.

It should be noted that the effects of the resonant sign-changing energy exchange in the SLs with sinusoidal miniband were discussed in Refs. [19,20,30,31] for harmonic and static fields and in Refs. [15,19] for the components of the biharmonic field (with and without the static component). Experimentally, the resonant interaction of the harmonic and static field was observed in Ref. [30] and the observation of ANC was reported in Ref. [32]. In Ref. [31], the used SLs were of poorer quality and these effects were not seen. However, in this work authors observed the theoretically predicted strong modification of the current-voltage characteristics of the SLs subject to strong terahertz fields. This strong modification of the current-voltage characteristics can be used for the terahertz field generation on the Bloch frequency and its harmonics in the regime of suppressed low-frequency domain instability (Gann effect). Such possibility for the SLs with sinusoidal miniband was discussed in Refs. [25,26,29].

Our analysis shows that SLs with non-sinusoidal minibands in the presence of muti-frequency fields exhibit richer variety of resonant field interaction effects than SLs with sinusoidal minibands in the presence of harmonic fields. In particular, in the resonant region with $\nu\Omega_c > \sum_{\alpha=1}^{M}\mu_\alpha^{(0)}\omega_\alpha$ , it is possible to have the situation when the energy not the static field only but that of many harmonics as well is transferred into some specific harmonics. Moreover, in this case in the contrast to the SLs with sinusoidal miniband, it happens at smaller static fields ($\nu >$ 1!), where the NDC is absent. Furthermore, in the SLs with non-sinusoidal miniband, the region of the NDC can be shifted toward the stronger fields even in the absence of the high-frequency component [33]. These properties of the fields interaction can be useful for the development of terahertz sources based on the SLs with non-sinusoidal minibands.

## 6. Peculiarities of the interaction between the fields with multiple frequencies in SLs

Let us examine the case of a periodic multi-frequency field as exemplified by a biharmonic field with multiple frequencies $\omega_2 = n_0\omega_1$, $n_0 = 2,3,\ldots$ , $\delta_1 = 0$, $\delta_2 \equiv \delta$, and $E_c \neq 0$. According to Eqs. (25,32), under this field, the steady current is the SL has the form

$$j(t) = \sum_{\nu=1}^{N}\sum_{\lambda_{1,2}=-\infty}^{\infty} j^{(\nu)}_{\lambda_1\omega_1,\lambda_2\omega_2}\exp[-i\lambda_1(\omega_1 t-\delta_1)-i\lambda_2(\omega_2 t-\delta_2)] + \text{c.c.}, \qquad (53)$$

where

$$j^{(\nu)}_{\lambda_1\omega_1,\lambda_2\omega_2} = \frac{i}{2}j_{0\nu}\sum_{\mu_{1,2}=-\infty}^{\infty}\frac{J_{\mu_1}(\nu g_1)J_{\mu_1+\lambda_1}(\nu g_1)J_{\mu_2}(\nu g_2)J_{\mu_2+\lambda_2}(\nu g_2)}{1+i(\nu\Omega_c+\mu_1\omega_1+\mu_2\omega_2)\tau}. \qquad (54)$$

It can be seen from Eqs. (32,54) that the effects of the resonant fields interaction, discussed in the previous section, can be applicable for the periodic multi-frequency fields. Moreover, in this case these effects are even more pronounced because in addition to the nonsynchronous conductivities of Eqs. (48),(49), the synchronous partial conductivities have the similar properties as well [20,34].

It follows from Eqs. (53,54) that the static component is given by

$$j_c = \sum_{\nu=1}^{M} j_c^{(\nu)}, \quad j_c^{(\nu)} = \sigma_c^{(\nu)}(\Omega_c,\Omega_1,\Omega_2,\omega_1,\omega_2)E_c + \Delta j_c^{(\nu)}(\Omega_c,\Omega_1,\Omega_2,\omega_1,\omega_2,\delta), \tag{55}$$

where nonsynchronous partial conductivities $\sigma_c^{(\nu)}(\Omega_c,\Omega_1,\Omega_2,\omega_1,\omega_2)$ can be found from Eq. (48) with $M$ = 2 and the partial rectification current (dependent on the phase relations of the fields) has the form

$$\Delta j_c^{(\nu)}(\Omega_c,\Omega_1,\Omega_2,\omega_1,\omega_2,\delta) = j_{0\nu} \sum_{\substack{\lambda=-\infty \\ \lambda\neq 0}}^{\infty} \sum_{\mu_{1,2}=-\infty}^{\infty} \frac{J_{\mu_1}(\nu g_1) J_{\mu_2}(\nu g_2) J_{\mu_1 - n_0\lambda}(\nu g_1) J_{\mu_2+\lambda}(\nu g_2)}{1 + [\nu\Omega_c + (\mu_1 + n_0\mu_2)\omega_1]^2 \tau^2}$$
$$\times \{-\sin(\lambda\delta) + [\nu\Omega_c + (\mu_1 + n_0\mu_2)\omega_1]\tau \cdot \cos(\lambda\delta)\} \tag{56}$$

In the general case, for the odd $n_0$ the rectification current $\Delta j_c^{(\nu)}$ is asymmetric with respect to $E_c$, and for even $n_0$ it does not have a specific symmetry. If the second component of the biharmonic field is weak ($g_2 << 1$), then the partial rectification current, Eq. (56), is given by

$$\Delta j_c^{(\nu)} = \frac{\sigma_{0\nu}}{n_0\omega_1} \sum_{\mu=-\infty}^{\infty} \frac{\nu\Omega_c + \mu\omega_1}{1 + (\nu\Omega_c + \mu\omega_1)^2\tau^2} J_\mu(\nu g_1)\left[J_{\mu-n_0}(\nu g_1) - J_{\mu+n_0}(\nu g_1)\right]E_2 \cos(\delta). \tag{57}$$

For even $n_0$ this current is symmetric with respect to $E_c$. The rectification current, Eq. (57), in the SLs with the sinusoidal miniband was studied in Refs. [17,19] for arbitrary $n_0$ and $\delta$ and in Refs. [27,35,36] for the presence of the biharmonic field containing the fundamental and second harmonics with arbitrary amplitudes $E_{1,2}$ and phase $\delta$. In the present work we examine the case of odd $n_0$ when there is no steady static current in the absence of the static field and arbitrary amplitudes $E_{1,2}$ and phase $\delta$.

It can be shown that in the biharmonic field, the real part of nonsynchronous linear conductivity on the third frequency $\omega_3$ ($g_3 << 1$), incommensurable with frequencies $\omega_{1,2}$, $\omega_3 \neq n_1\omega_1 + n_2\omega_2$, $n_{1,2} = 0,\pm1,\pm2,\ldots$, is given by (in the $\tau$-approximation)

$$\mathrm{Re}\left[\sigma_{\omega_3}^{(\nu)}(\Omega_c,\Omega_1,\Omega_2,\omega_1,\omega_2,\omega_3,\delta)\right] = \frac{1}{2}\left[\left(1 + \frac{\nu\Omega_c}{\omega_3}\right)\sigma_c^{(\nu)}\left(\Omega_c + \frac{\omega_3}{\nu},\Omega_1,\Omega_2,\omega_1,\omega_2,\delta\right)\right.$$
$$\left. + \left(1 - \frac{\nu\Omega_c}{\omega_3}\right)\sigma_c^{(\nu)}\left(\Omega_c - \frac{\omega_3}{\nu},\Omega_1,\Omega_2,\omega_1,\omega_2,\delta\right)\right]. \tag{58}$$

In particular, it follows from Eq. (58) that

$$\mathrm{Re}\left[\sigma_{\omega_3}^{(1)}(0,\Omega_1,\Omega_2,\omega_1,\omega_2,\omega_3,\delta)\right] = \sigma_c(\Omega_c = \omega_3,\Omega_1,\Omega_2,\omega_1,\omega_2,\delta). \tag{59}$$

Eqs. (58,59) remain valid for any periodic field, not for the biharmonic field only.

The expressions for the alternating current harmonics are too cumbersome and not presented here. Instead, for numerical calculations we use the following integral expression for the partial current obtained from Eqs. (21,24,25,31,33):

$$j_\nu(t)=\left\{\tilde{j}_{0\nu}\sin\left[\frac{e\nu d}{\hbar}\int_{t_0}^{t}E(t_2)dt_2\right]+ne\tilde{V}_{0\nu}\cos\left[\frac{e\nu d}{\hbar}\int_{t_0}^{t}E(t_2)dt_2\right]\right\}\exp\left(-\frac{t-t_0}{\tau}\right)$$
$$+j_{0\nu}\int_{t_0}^{t}\frac{dt_1}{\tau}\exp\left(-\frac{t-t_1}{\tau}\right)\sin\left[\frac{e\nu d}{\hbar}\int_{t_1}^{t}E(t_2)dt_2\right]. \qquad (60)$$

As in the case of *non-periodic* biharmonic field, for the periodic field, the current-voltage characteristics contain the regions with NDC and ANC. As an example, we show in Fig. 3 the regions of negative nonsynchronous static conductivity $\sigma_c^{(1)}(\Omega_c,\Omega_1,\Omega_2,\omega_1,\omega_2)$ on the $E_1$-$E_c$ plane in the SL with sinusoidal miniband in the periodic biharmonic field with at $n_0 = 3$, $\delta = \pi$ (a), $\delta = \pi/2$ (b), and $\omega_1\tau = 10$. Coupled values of amplitudes $E_{1,2}$ correspond to the electron DL and the SIT of the SL in this field (curves (1) in Fig. 1). According to Eq. (59), in the same regions the high-frequency field on the frequency $\omega_3$ (non-commensurable with frequencies $\omega_{1,2}$) is linearly non-parametrically amplified at $E_c = 0$. Because the coupling between the field amplitudes $g_1$ and $g_2$ is strongly nonlinear in the DL region at $\delta = \pi/2$ (see Fig. 1(b)), we divide Fig. 3 into two parts: the left panel corresponds to the horizontal part of the curve (1) and shows the dependence of $\Omega_c$ on $g_1$ at $g_2 \approx 2.4$, and the right panel corresponds to the vertical part of the curve (1) and shows the dependence of $\Omega_c$ on $g_2$ at $g_1 \approx 2.4$. The boundary curves of the ANC regions where the static conductivity of the SL is zero ($\sigma_c^{(1)}(\Omega_c,\Omega_1,\Omega_2,\omega_1,\omega_2)=0$), correspond to the SL states stable (solid lines) and unstable (dashed lines) with respect to quazistatic field perturbations. Here we do not take into account parametric instabilities, which are weakly pronounced in the SLs under SIT conditions due to the smallness of corresponding nonlinear currents, especially at low electron density.

It is evident that the SIT regions are located inside the regions of ANC and NDC (similar to the case of the harmonic field [14,17,19]). Consequently, the SIT states at the biharmonic field are unstable with respect both to the significant static field generations and to the excitations of the high-frequency (plasma, in particular) oscillations. The former is the main mechanism of the SIT vanishing at low electron concentrations. The steady values of the generated static fields in the circuit open with respect to the static current satisfy the equation

$$\sigma_c(\Omega_c,\Omega_1,\Omega_2,\omega_1,\omega_2)E_c+\Delta j_c(\Omega_c,\Omega_1,\Omega_2,\omega_1,\omega_2,\delta)=0. \qquad (61)$$

One can see from Eqs. (48), (56), and (61) that the steady states of $E_c$ at the given alternating internal field do not depend on the electron concentration. However, the electron density determines the character of the transient processes and the final state, if several of them are possible. The existence of the nonlinear rectification current (61) in biharmonic field leads to different scenario of the static field generation than that of Ref. [19]. In particular, the values of the generated static fields, given by Eq. (61), do not coincide with the boundary curves of the ANC regions, given by

$\sigma_c(\Omega_c, \Omega_1, \Omega_2, \omega_1, \omega_2) = 0$. However, the numerical solutions of Eq. (61) are almost the same as given in Fig. 3 for the same parameters. The reason for this is following. As was shown above, in the high-frequency range ($\omega_1\tau >> 1$) significant energy exchange between fields occurs in narrow resonant regions given by relations $|\nu\Omega_c - n\omega_1| \approx \tau^{-1}$, $n = 1,2,...$ , where both conductivities and currents are large. In these regions the rectification current (56) also has the large values. Consequently, the solutions of Eq. (61) describing the compensation of the static currents should lay in these regions as well. As far as the low-frequency range ($\omega_1\tau \sim 1$) is concerned, the values of the generated static fields differ significantly from the values at the boundaries of the ANC regions [26]. It should be emphasized that at the given biharmonic field (similar to the case of the harmonic field [19]), several stable steady "currentless" ($j_c = 0$) states can exist with different values of $E_c$. Moreover, there is a hysterisis in the dependencies of $E_c$ on the fields $g_1$ and $g_2$. It is also evident from Fig. 3 that the minimal static field can be different in the steady state depending on the relation between biharmonic field components. In the case of $\delta = \pi/2$, there is pronounced step-like transition from the generation of $\Omega_c = \omega_1$ (at $g_1 >> g_2$) to the generation of $\Omega_c = 3\omega_1$ ($g_2 >> g_1$). In the regions where the components are comparable, the minimal static field is determined by the frequency $2\omega_1$ which is absent in the electric field but present in the nonlinear *ac* BOs.

The generated static field changes the high-frequency conductivities of the SL (including their sign), which leads to the energy redistribution between the field components. As a result, the magnitude and the spectral constitution of the internal field can be changed significantly. This is illustrated in Fig. 4: increasing of the static field increases the region and the magnitude of the negative field absorption on the frequency $\omega_2 = 3\omega_1$. In turn, the amplitude increase of this component can lead to the decrease of the ANC value and even to its vanishing (see Fig. 3). As a result, the increase of the static field can be replaced by its decrease. Moreover, the parametric processes become to be significant when the SIT state decays. The competition between all these effects can prolong the transient process of the SIT vanishing and even lead to its partial stabilization or to stochastic oscillations (at high electron concentrations).

In the general case, the states to which the SL in an external periodic field is transferred from the SIT state are nonlinear oscillations of a current and an internal field, where the static component can exist as well. Such states can be more than one which, in particular, can be seen in the multiple values and even hysteresis in the dependencies of the static field magnitude and the current (and emission) spectrum on the external field parameters. At least, five scenarios of SL behavior are possible: (i) SIT collapse accompanying by the static field generation; (ii) SIT collapse accompanying by the significant amplifications of one or several components of the periodic field and the corresponding currents; (iii) partial stabilization of the SIT caused by the energy exchange between the components of the internal field (including parametric processes of the modes decomposition and recomposition) leading to a suppression of both the static field generation and the amplification of the high-frequency harmonics with the combinational frequencies of the external field. Such stabilization can be stimulated by an addition of corresponding weak component to the external field. The following processes are also possible: (iv) SIT collapse accompanying by the amplification of the harmonics and subharmonics of the external field; and (v) SIT collapse accompanying by the generation and amplification of harmonics with frequencies $\omega_r$ close to external field harmonics $n\omega$, so $|n\omega - \omega_r| \sim \tau^{-1}$, $n = 1,2,...$ In the SLs with low electron density ($(\omega_0/\omega)^2 << 1$), the first three scenarios can take place. To analyze such possibilities, we perform the numerical simulations with results showed below.

### 7. SL current oscillations in an external field

We describe the time evolution of a SL current, the electron energy, and an internal SL field in an external field by the kinetic equation, Eq. (23), and the continuity equation for the full current:

$$\varepsilon_0 \frac{dE(t)}{dt} + 4\pi j(t) + \frac{\varepsilon_0 E(t)}{RC_S} = 4\pi j_e(t),$$

(62)

where $R$ is the resistance shunting the SL (for simplicity it can be chosen to be infinitely large), $\varepsilon_0$ is the dielectric constant of the SL without electron contribution, $C_s$ is the linear capacitance of the SL, and $j_e(t)$ is the external current density depending of the details of the SL connection to an external circuit. We restrict ourselves by the case of the circuit with the given external field $E_e(t)$ (circuit disconnected with respect to the direct current). In this situation,

$$j_e(t) = \frac{\varepsilon_e}{4\pi} \frac{\partial E_e(t)}{\partial t}, \quad (63)$$

where $\varepsilon_e$ is the permittivity of the external medium. Other possibilities of the SL inclusion to the external circuit were discussed in Refs. [19,22,23].

Let us assume that the given external field contains $N_e$ harmonics:

$$E_e(t) = \sum_{\alpha=1}^{Ne} E_\alpha^{(e)} \cos(\omega_\alpha t - \delta_\alpha), \quad \delta_\alpha = \text{const.} \quad (64)$$

We introduce the dimensionless variables: $\tilde{t} = t\omega_1 - \delta_1, g(t) = \Omega(t)/\omega_1$, $\tilde{V}_\alpha^{(e)} = eE_\alpha{}^{(e)} d\varepsilon_e/\hbar\omega_1\varepsilon_0$, $\tilde{V}_e(\tilde{t}) = eE^{(e)}(t) d\varepsilon_e/\hbar\omega_1\varepsilon_0$, $n_\alpha = \omega_\alpha/\omega_1$, $\varphi_\alpha = \delta_\alpha - n_\alpha\delta_1$, and $\tilde{w}_\nu = (\omega_{0\nu}/\omega_1)^2$, where $\omega_{0\nu}^2 = \frac{4\pi e d}{\hbar\varepsilon_0}\nu j_{0\nu} = \frac{4\pi}{\varepsilon_0\tau}\sigma_{0\nu} = \frac{4\pi\rho e^2}{\varepsilon_0}\left(\frac{\nu d}{\hbar}\right)^2\left(\frac{\Delta_\nu}{2} - \langle\varepsilon_\nu\rangle_0\right)$ is the partial plasma frequency squared. Correspondingly, from Eqs. (18),(62),(63) we obtain the closed set of equations for the electric current, $j(t)$, the averaged longitudinal electron energy, $\langle\varepsilon_3\rangle$, and the electric field in the SL, $E(t)$, in the dimensionless variables:

$$\omega_1\tau\frac{d\Phi_\nu(\tilde{t})}{d\tilde{t}} + \left[1 + i\nu\omega_1\tau g(\tilde{t})\right]\Phi_\nu(\tilde{t}) = 1, \tag{65}$$

$$\frac{dg(\tilde{t})}{d\tilde{t}} = \sum_{\nu=1}^{N}\nu^{-1}\left(\frac{\omega_{0\nu}}{\omega_1}\right)^2 \operatorname{Im}\Phi_\nu(\tilde{t}) + \frac{d\tilde{V}_e}{d\tilde{t}}, \tag{66}$$

and

$$\tilde{V}_e(\tilde{t}) = \sum_{\alpha=1}^{N_e}\tilde{V}_\alpha^{(e)}\cos(n_\alpha\tilde{t} - \varphi_\alpha). \tag{67}$$

Sometimes, instead of differential equations (65) and (66), it is convenient to use the integral-differential equation followed from them:

$$\frac{dg(\tilde{t})}{d\tilde{t}} + \sum_{\nu=1}^{N}\nu^{-1}\left(\frac{\omega_{0\nu}}{\omega}\right)^2 \tilde{j}_\nu(\tilde{t}) = \frac{d\tilde{V}_e}{d\tilde{t}}, \tag{68}$$

where, according to Eq. (60),

$$\begin{aligned}\tilde{j}_\nu(\tilde{t}) \equiv \frac{j_\nu(\tilde{t})}{j_{0\nu}} &= \left\{\frac{\tilde{j}_{0\nu}}{j_{0\nu}}\sin\left[\nu\int_{\tilde{t}_0}^{\tilde{t}} g(\tilde{t}_1)d\tilde{t}_1\right] + \frac{\rho e\tilde{V}_{0\nu}}{j_{0\nu}}\cos\left[\nu\int_{\tilde{t}_0}^{\tilde{t}} g(\tilde{t}_1)d\tilde{t}_1\right]\right\}\exp\left(-\frac{\tilde{t}-\tilde{t}_0}{\omega_1\tau}\right)\\ &+ \int_{\tilde{t}_0}^{\tilde{t}}\frac{d\tilde{t}_1}{\omega_1\tau}\exp\left(-\frac{\tilde{t}-\tilde{t}_1}{\omega_1\tau}\right)\sin\left[\nu\int_{\tilde{t}_1}^{\tilde{t}} g(\tilde{t}_2)d\tilde{t}_2\right].\end{aligned} \tag{69}$$

Based on these equations, we examine the above-mentioned scenarios of the SIT collapse/stabilization in the SLs with relatively low electron density ($(\omega_0/\omega_1)^2 = 0.05$). In Figs. 5 and 6 we show the time evolutions of the current and the field averaged over the period in the SL containing one, two, or three partial minibands with $\Delta_n = (-1)^{n+1}n^{-2}\Delta_1$, $n = 1,2,3$ (which corresponds to the quasiparabolic miniband approximation) in the presence of the periodic external field. In Fig. 5 the external field has a form of the set of rectangular pulses with $\Omega_c = 4\omega_1$ (for the numerical solution we used the approximation of this signal by the function $g(t) = g_1\tanh(10\cdot\sin\omega t)$), while in Fig. 6 the biharmonic field of the form

$$\tilde{V} = \tilde{V}_1\sin(\omega_1 t) + \tilde{V}_2\sin(3\omega_1 t - \pi/2) \tag{70}$$

with $\tilde{V}_1 = 2.39$ and $\tilde{V}_2 = 1.9$ is applied. In both cases we have $\omega_1\tau = 10$. One can see from these figures that the SIT state is formed for the time of the order of several $\tau$ (in the beginning of the processes the current corresponds to the *ac* BOs and contains harmonics with frequencies $n\omega_1$, $n = 0,1,2,\ldots$), exists during several hundred of the field periods (in this situation, the current spectrum contains only the odd frequency, $(2n+1)\omega_1$), and, thereafter, collapses with the generation of the static field with $g_c = \pm 1$ and the current containing both odd and even frequencies (except the zero-th one). As was mentioned, in the pulsed external field the DL and SIT occur in SLs with arbitrary miniband dispersion law. However, the dispersion law affects the lifetime of the SIT and the character of its collapse, which is evident from Figs. 5 and 6.

The time evolution (corresponding to the second scenario) of the current and its spectral constitution in the SL with the sinusoidal miniband in the biharmonic field

$$\tilde{V} = \tilde{V}_1 \sin(\omega_1 t) + \tilde{V}_2 \sin(3\omega_1 t - \pi) \tag{71}$$

with $\tilde{V}_1 = 2.73$, $\tilde{V}_2 = 0.56$, and $\omega_1\tau = 10$ are shown in Fig. 7. It is evident that initially there is significant alternating current and the static field generation occurs. With its growth, however, the current and internal field components on the $3\omega_1$ frequency increase as well, that changes the direction of the energy exchange between the static and high-frequency components, so the static field vanishes and the alternating current becomes quite small ($j/j_0 \approx 0.1$). Thereafter, as a result of continuing energy exchange, the alternating current starts to grow and reaches a significant value ($j/j_0 \approx 0.35$) and the amplitude of the current component on the $3\omega_1$ frequency can reach the values larger than that of the fundamental frequency. As a result, the SIT collapses and the SL has a steady state with a significant alternating current but without the static component.

Fig. 8 demonstrates the relative stabilization of the transparency in the SL with sinusoidal miniband by the higher (third and fifth) field and current harmonics in the presence of the biharmonic external field of Eq. (71) with $\tilde{V}_1 = 4.97$, $\tilde{V}_2 = 0.48$, and $\omega_1\tau = 10$. This corresponds to the third scenario. The alternating current, significantly grown on the first field periods ($j/j_0 \approx 0.7$), rapidly decay to relatively small values ($j/j_0 \approx 0.1$) and remain the same as time elapses. This is caused by the fact that the electron gas becomes strongly heated and fills the Brillouin miniband almost homogeneously.

## 8. Conclusions

In conclusions, we have studied the electromagnetic transparency of semiconductor superlattices (SLs) with various miniband dispersion laws in multi-frequency periodic and non-periodic electric field both with and without the static component. We have determined the processes of its formation, collapse, and stabilization. We have shown that in the general case the SL transparency is possible only in the presence of the multi-frequency field with the number of components equal to the number of the Fourier-harmonics of the miniband dispersion law. However, in some specific cases the number of the field components can be twice smaller. If the *ac* Bloch oscillations (*ac* BOs) do not contain low-frequency harmonics, then in the constant relaxation time ($\tau$-) approximation (and in this approximation only!) the self-induced transparency (SIT) occurs at the same fields as the dynamical localization (DL) and the quasienergetic miniband collapse. In this case, in the $\tau$-approximation the DL is the reason for the SIT formation occurring in the time of the order of $\tau$ as a result of electron scattering. In the presence of a multi-frequency non-periodic field or a periodic field containing the static component with the Bloch frequency incommensurable with the field frequency, the *ac* BOs contain strongly decaying low-frequency harmonics which induce significant macroscopic currents in the SL on the field combinational frequencies. Correspondingly, there is an almost dissipationless resonant energy exchange between the field harmonics. In such fields, the DL does not lead to the SIT formation. However, at some specific values of the field component amplitudes, the energy exchange disappears and the SIT is formed. We have described five possible scenarios of the collapse or partial stabilization of the SIT state in the presence of external high-frequency fields. In the SLs with low electron density three of them can be seen: the SIT collapse caused by (a) the spontaneous static field generation or (b) significant amplification of one or several high-frequency field and current components, and (c) the partial SIT stabilization caused by the energy redistribution between the field components which suppresses both the static field generation and the harmonics amplification. We have presented the results of numerical modeling of the time evolution of the current in the SLs in the presence of an external biharmonic field, exhibiting such scenarios.

**Acknowledgements:**

The work of Yu.A.R. and J.Yu.R. is supported by the RFBR (grant No. 07-02-01126 ) and by the Program Basic Research of the Presidium of RAS No.27. L.M. was partially supported by the NSF NIRT, Grant No. ECS-0609146.

**Captions to the Figures:**

**Figure 1:** Regions of dynamical localization and self-induced transparency in the $g_1$-$g_2$ plane for the superlattices having either sinusoidal miniband (solid lines) or two partial minibands (crosses) in the presence of a biharmonic electric field with $\omega_2 = 3\omega_1$ and (a) $\delta = \pi$ and (b) $\delta = \pi/2$.

**Figure 2:** (Color online) Current-voltage characteristics (solid blue line) and real parts of high-frequency superlattice conductivities (arbitrary units), $\mathrm{Re}\,\sigma_{\omega_1}^{(1)}$ (dashed green line) and $\mathrm{Re}\,\sigma_{\omega_2}^{(1)}$ (dash-dotted red line), for $g_1 = g_2 = 2.405$, $\omega_2 = 1.3\omega_1$, and $\omega_1\tau = 10$.

**Figure 3:** Regions of absolute negative conductivity (marked by 1) in the $g_1$-$\Omega_c/\omega_1$ plane for $\omega_1\tau = 10$ and (a) $\delta = \pi$ and (b) $\delta = \pi/2$. The chosen value of $g_2$ corresponds to the SIT conditions (line 1 of Figure 1). Unstable states are shown by the dashed lines.

**Figure 4:** (Color online) Dissipative loss in the superlattice (arbitrary units) at the frequency $\omega_2 = 3\omega_1$ and the phase shift $\delta = \pi$ for various values of the static field. The chosen values of $g_2$ correspond to the SIT conditions (line 1 of Figure 1(b)).

**Figure 5:** (Color online) Time evolution of (a) current and (b) field averaged over the period for the superlattices containing one (solid blue line in the (b) panel), two (dashed green line), and three (dash-dotted red line) partial minibands with $\Delta_n = (-1)^{n+1}n^{-2}\Delta_1$, $\Delta_2 = -\Delta_1/4$, and $\Delta_3 = \Delta_1/9$ in the presence of the external field of the set of rectangular pulses with $\Omega_c = 4\omega_1$.

**Figure 6:** (Color online) The same dependencies as in Figure 5 for the biharmonic external field having the form $\tilde{V} = \tilde{V}_1 \sin(\omega_1 t) + \tilde{V}_2 \sin(3\omega_1 t - \pi/2)$ with $\tilde{V}_1 = 2.39$, $\tilde{V}_2 = 1.9$, and $\omega_1\tau = 10$.

**Figure 7:** Collapse of self-induced transparency with the third current harmonic amplification: time evolution of the normalized superlattice current in the presence of the biharmonic field $\tilde{V} = \tilde{V}_1 \sin(\omega_1 t) + \tilde{V}_2 \sin(3\omega_1 t - \pi)$ with $\tilde{V}_1 = 2.73$, $\tilde{V}_2 = 0.56$, and $\omega_1\tau = 10$. Inset: current spectral constitution after the collapse.

**Figure 8:** Stabilization of self-induced transparency with higher current harmonic amplification: time evolution of the normalized superlattice current in the presence of the external biharmonic field $\tilde{V} = \tilde{V}_1 \sin(\omega_1 t) + \tilde{V}_2 \sin(3\omega_1 t - \pi)$ with $\tilde{V}_1 = 4.97$, $\tilde{V}_2 = 0.48$, and $\omega_1\tau = 10$. Inset: current spectral constitution in the steady state.

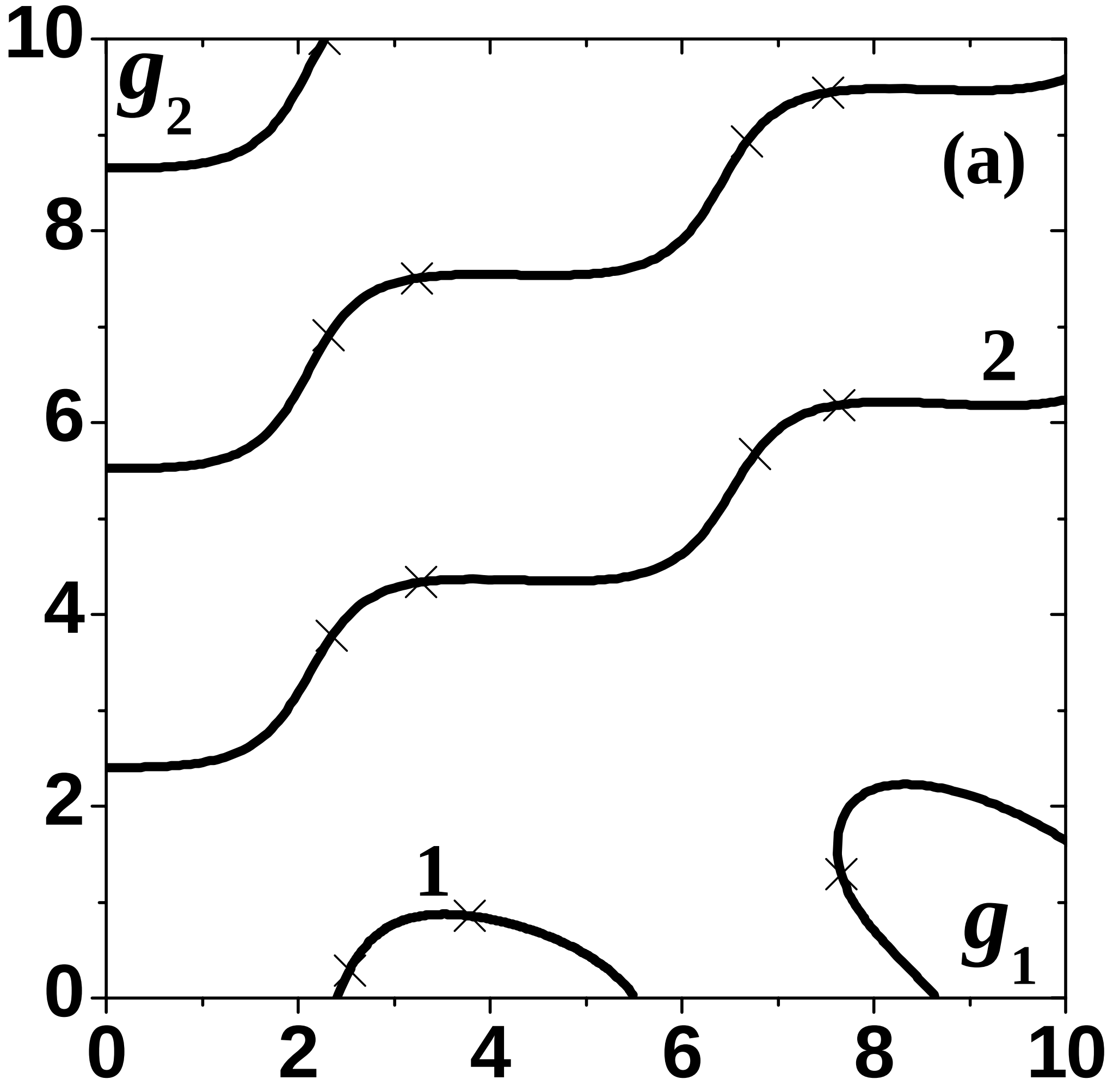

Figure 1(a).

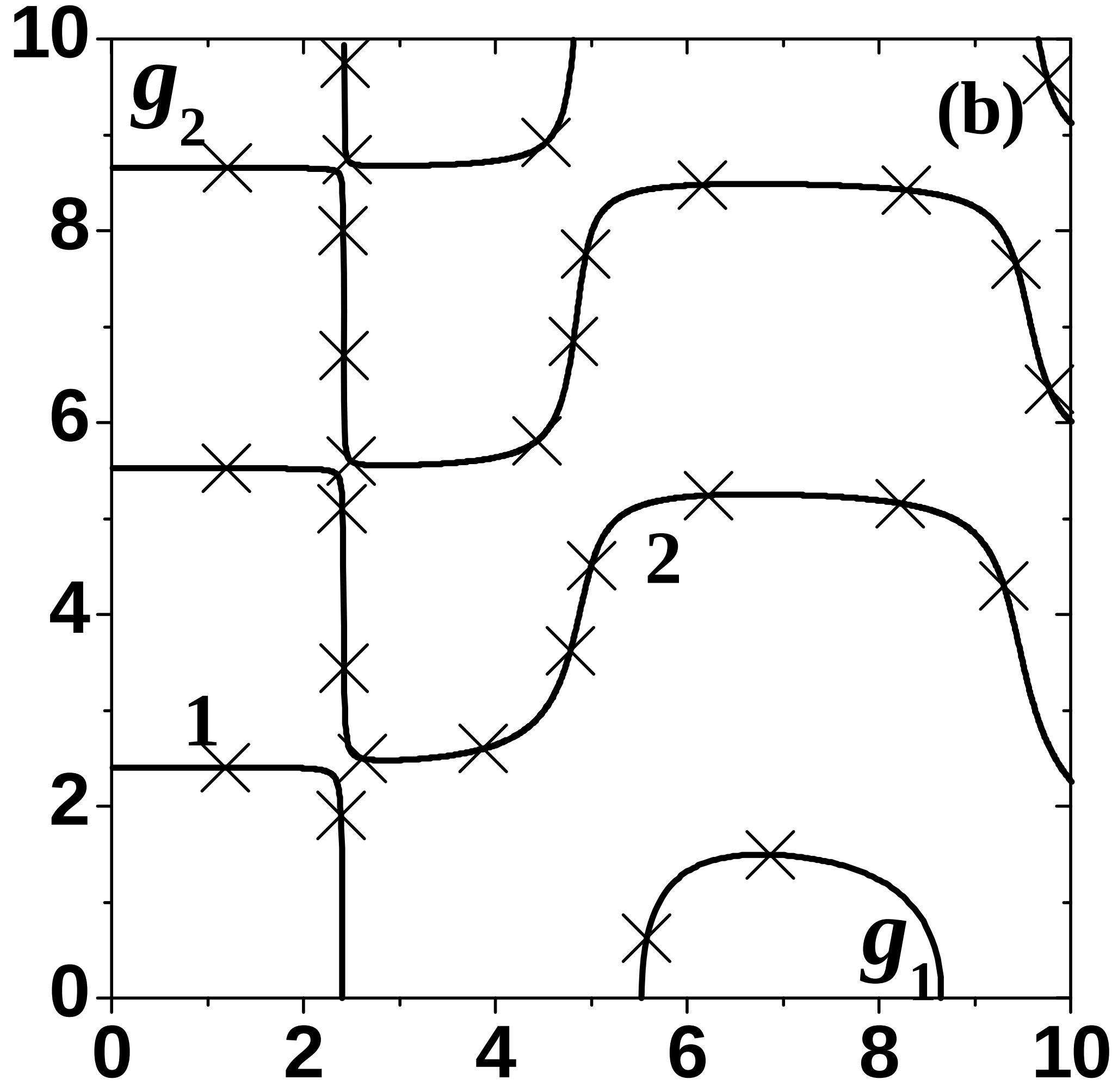

Figure 1(b).

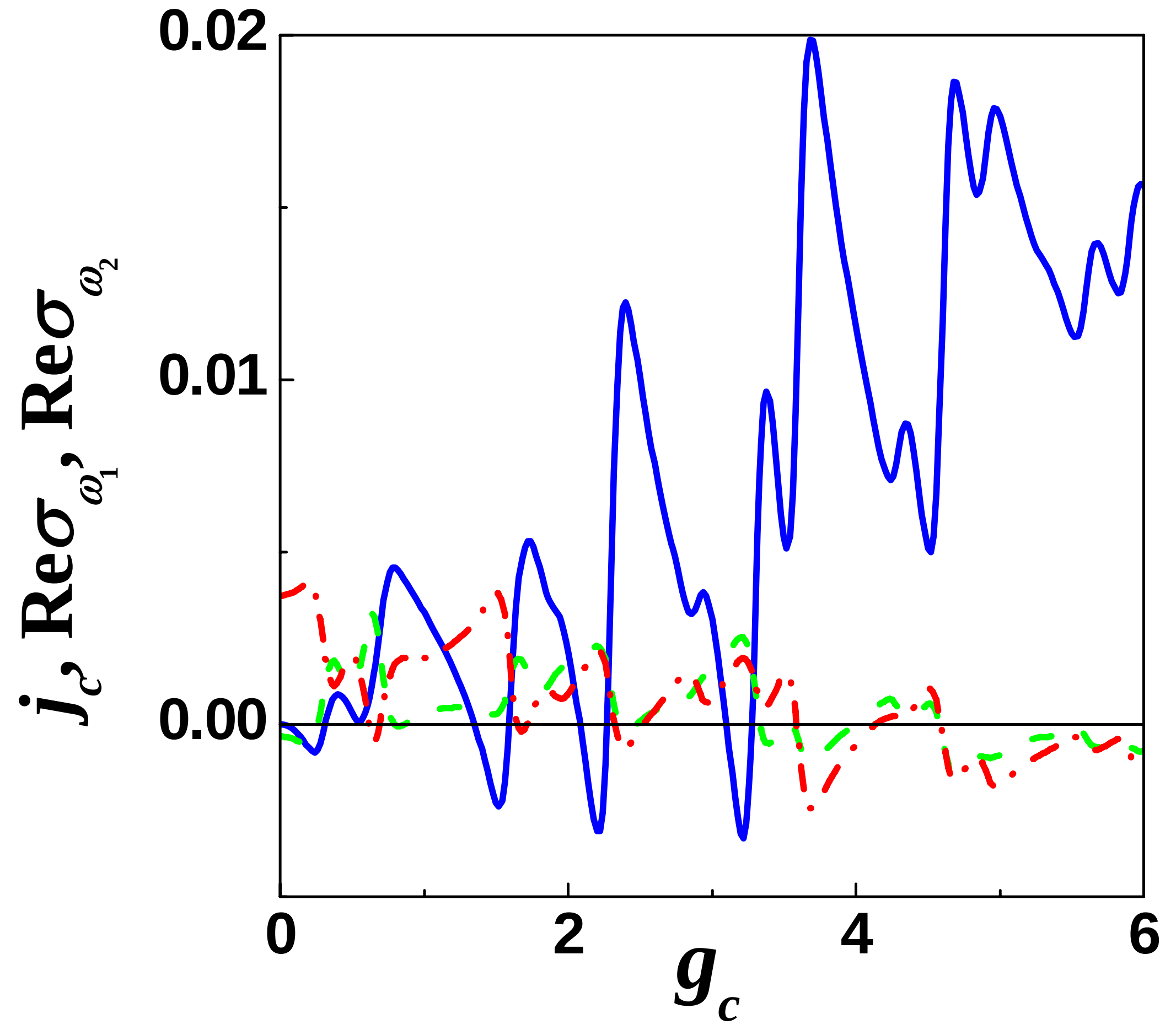

Figure 2.

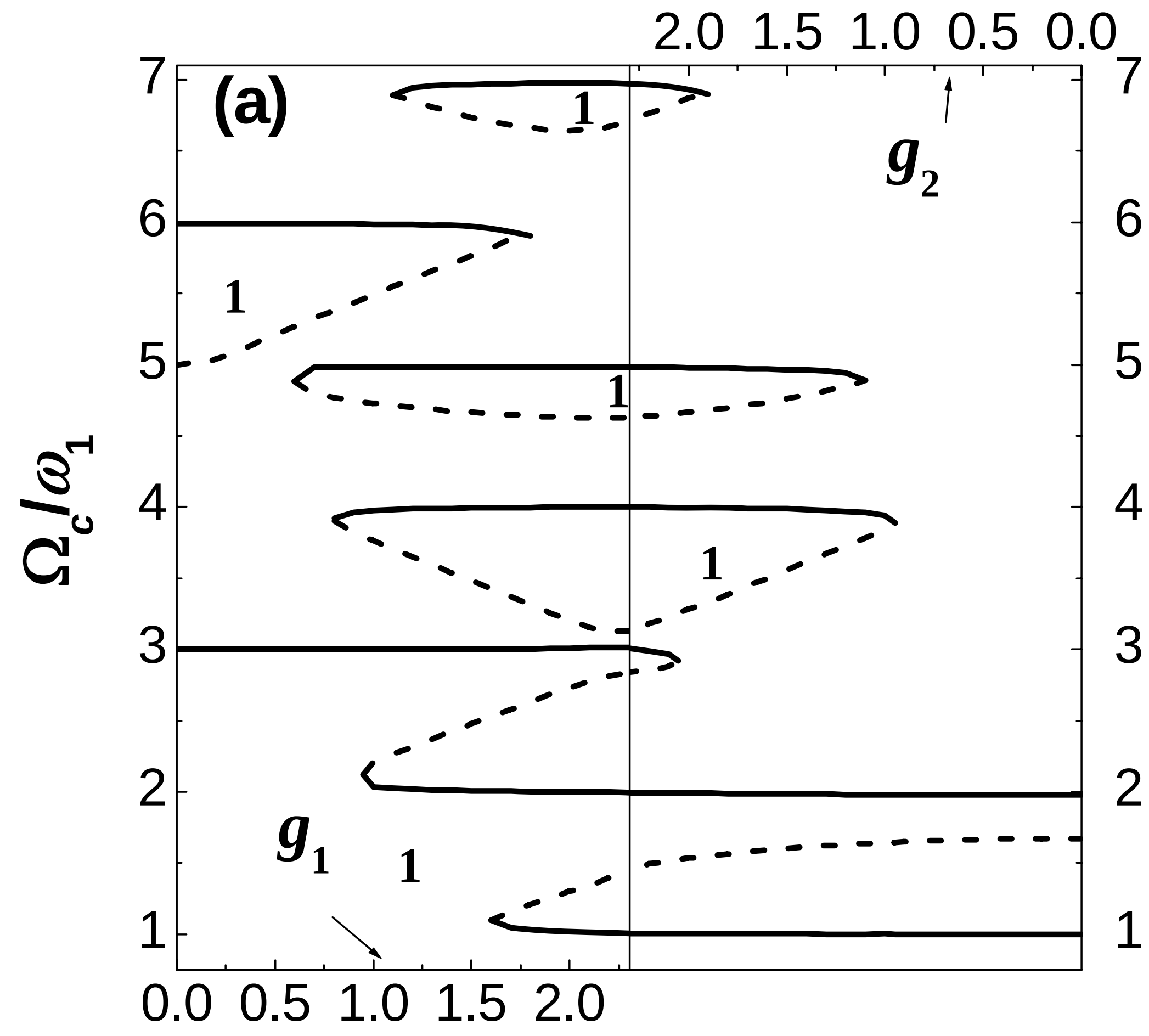

Figure 3(a).

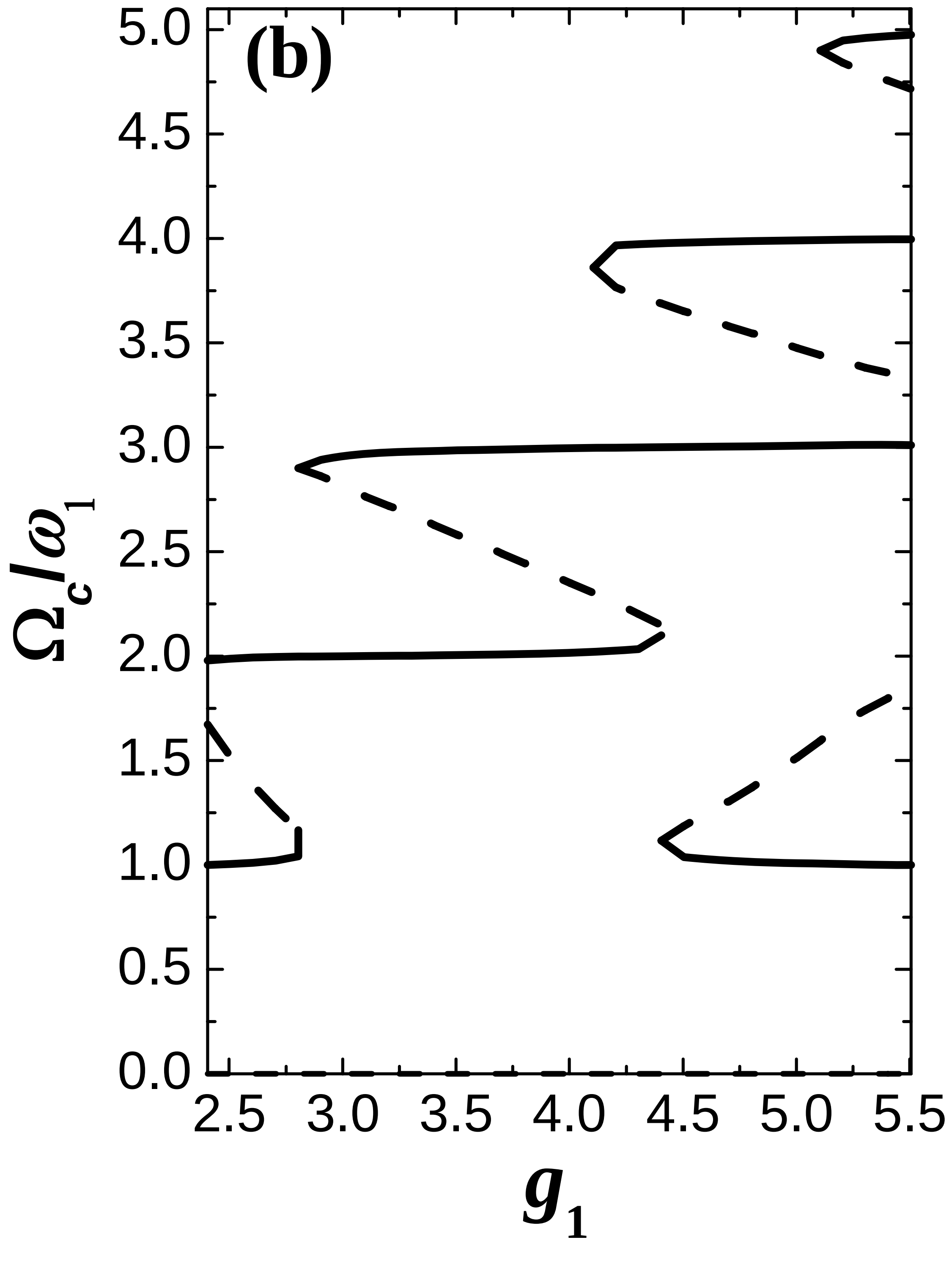

Figure 3(b).

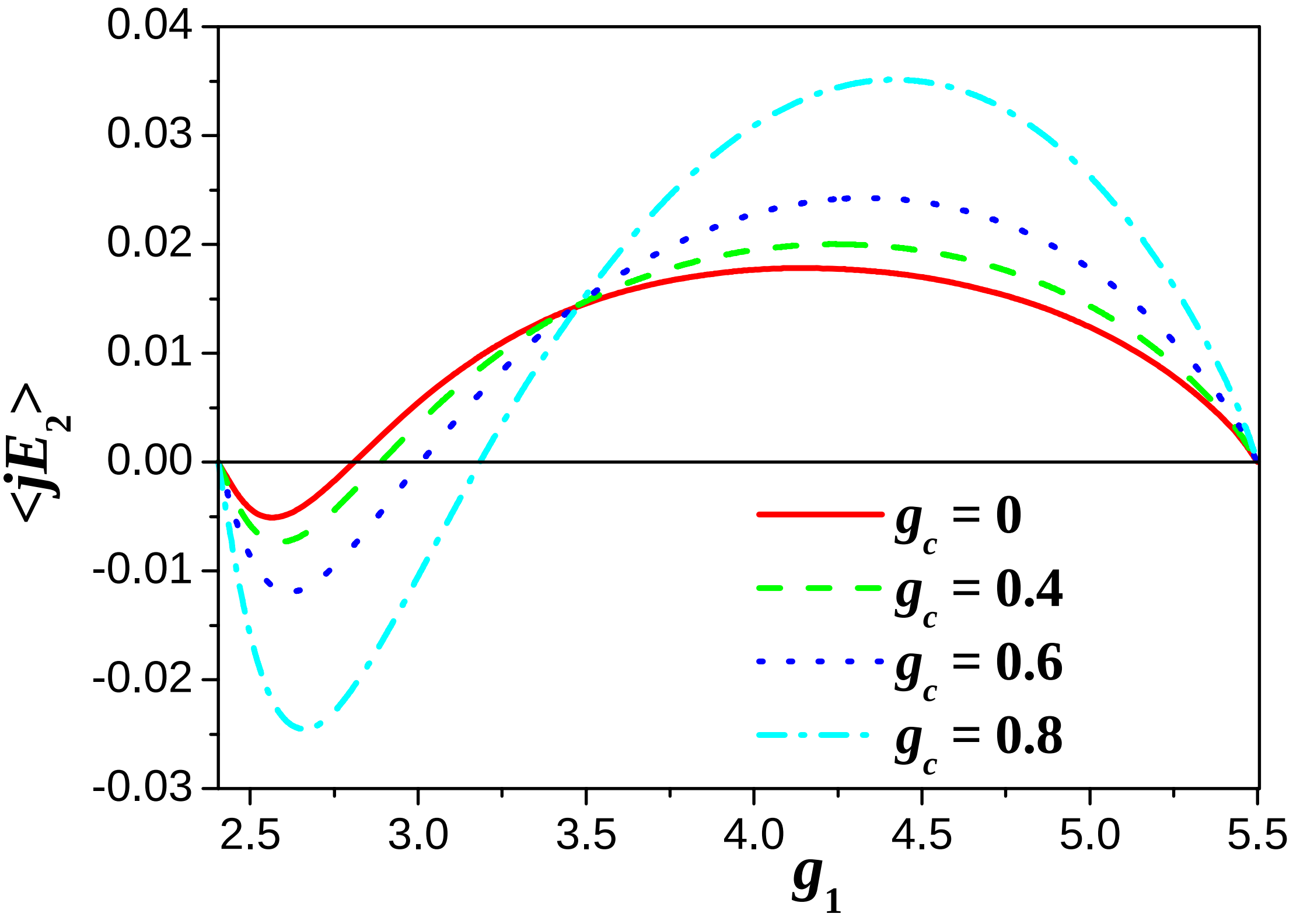

Figure 4.

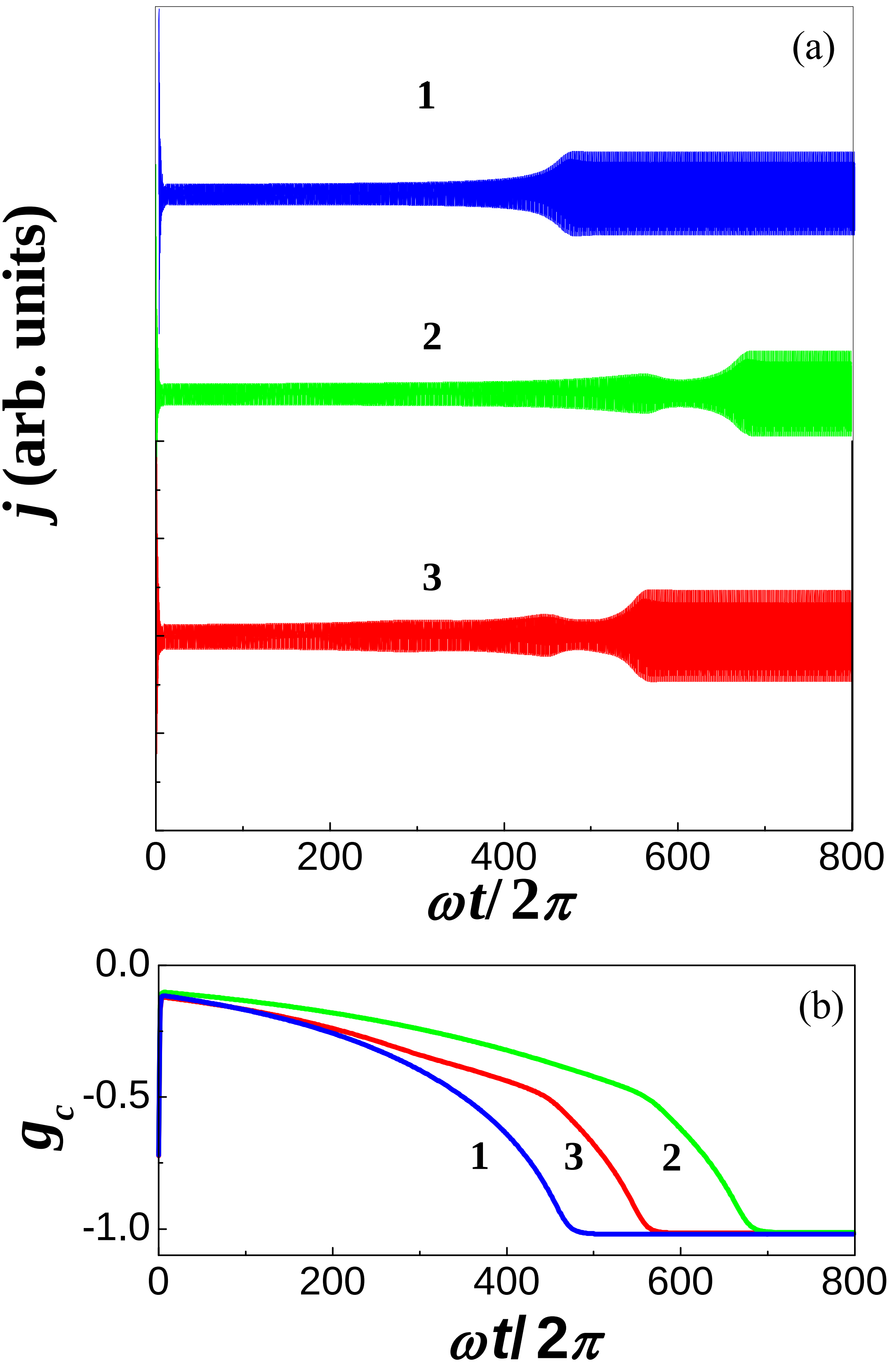

Figure 5.

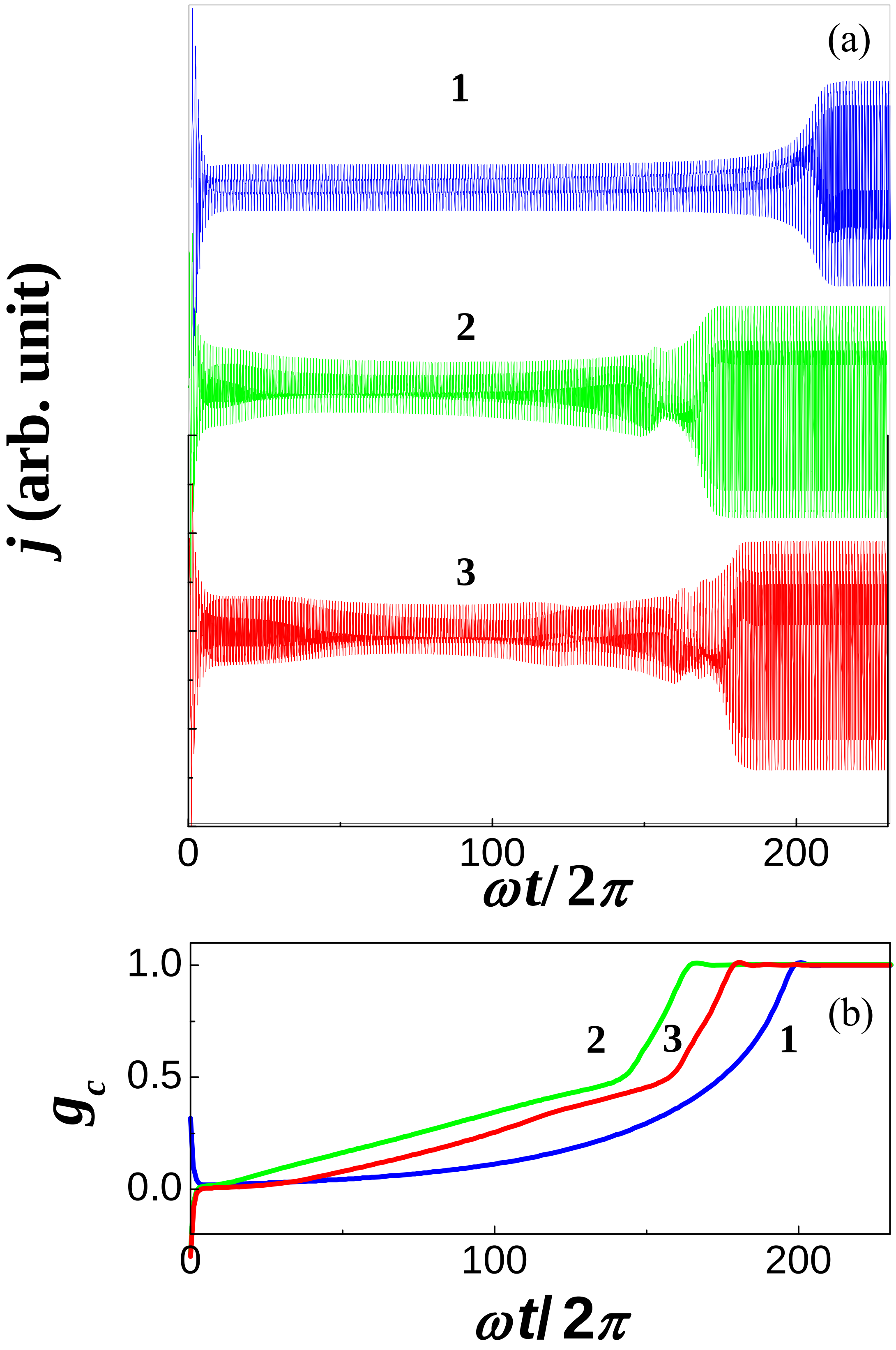

Figure 6.

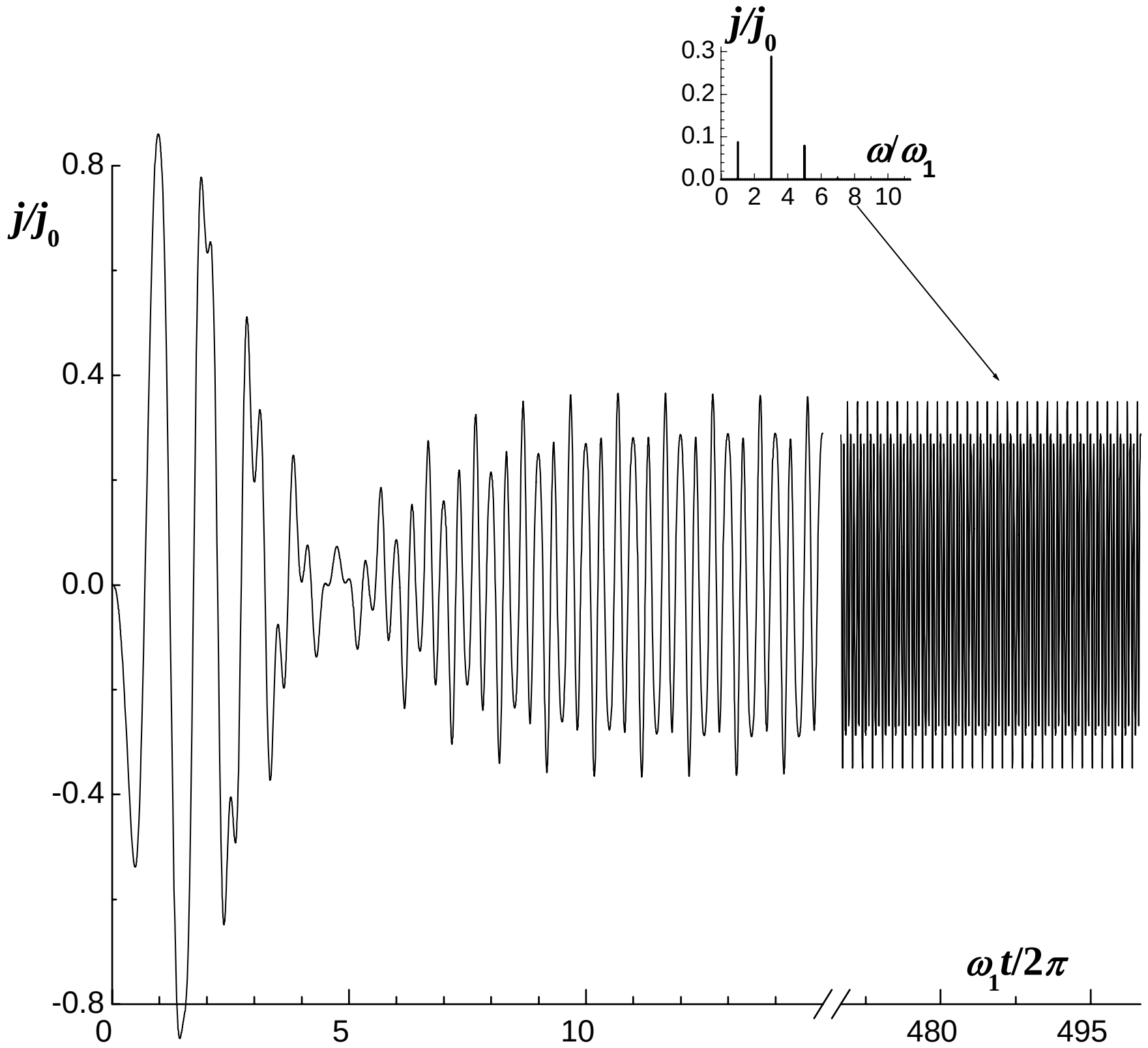

Figure 7.

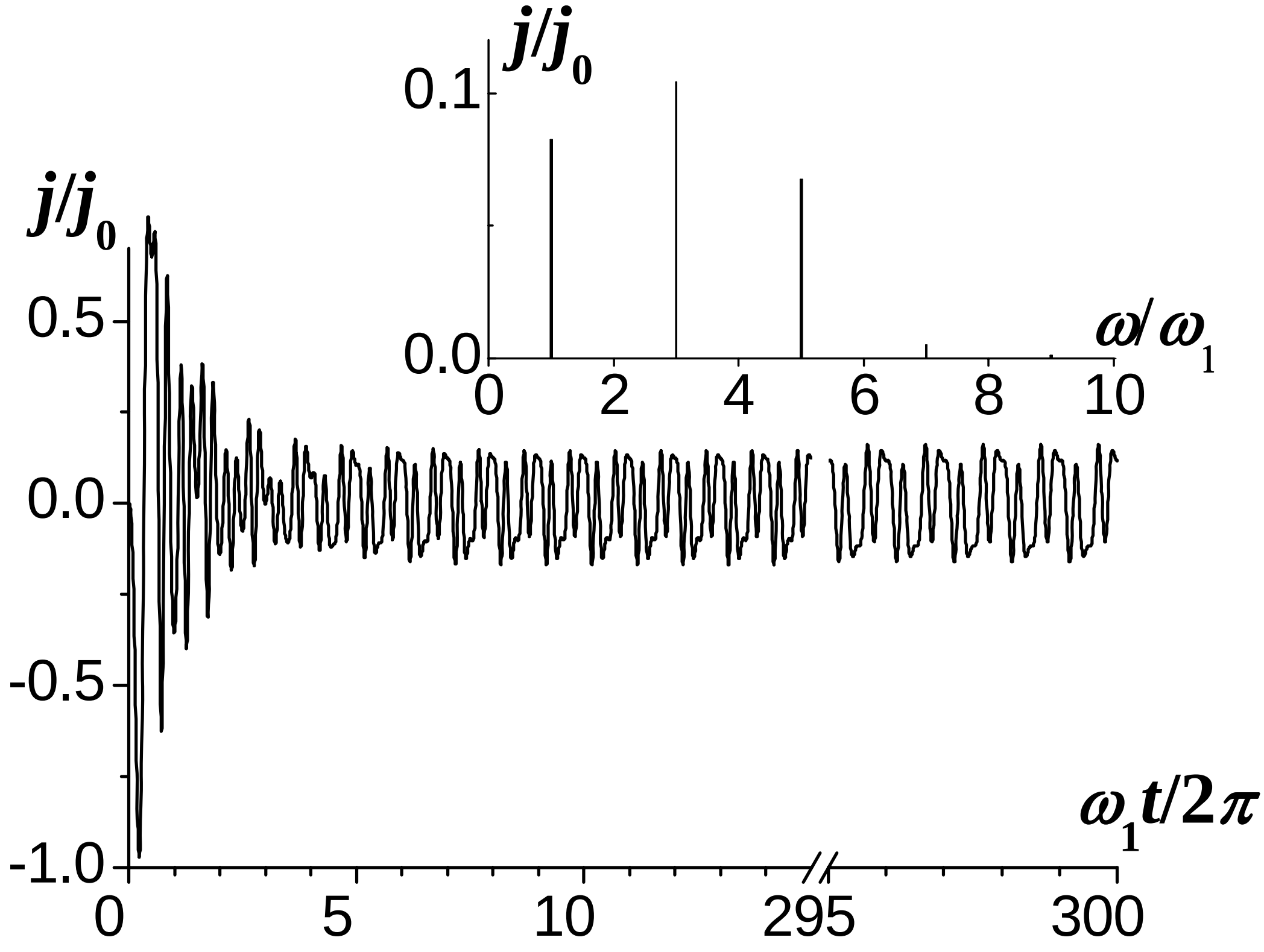

Figure 8.